\documentclass[pdflatex,sn-mathphys-ay]{sn-jnl}

\usepackage[utf8]{inputenc}
\usepackage{xcolor}
\usepackage{booktabs}
\usepackage{amsmath,amssymb}

\usepackage[capitalize]{cleveref}
\usepackage{graphicx}
\usepackage{tabularx}
\usepackage{rotating}

\newtheorem{prop}{Proposition}

\newcommand{\proba}[1]{p\left(#1\right)}
\newcommand{\likelihood}[1]{g\left(#1\right)}
\newcommand{\prior}[1]{\pi\left(#1\right)}
\newcommand{\posterior}[1]{f\left(#1\right)}
\newcommand{\abcposterior}[1]{\tilde{f}\left(#1\right)}
\newcommand{\kldiv}[2]{D_\mathrm{KL}\left(#1\;\Vert\;#2\right)}
\newcommand{\entropy}{H}
\newcommand{\epe}{\mathcal{H}}

\newcommand{\diff}[1]{\mathrm{d}#1\,}

\makeatletter
\def\E{\@ifnextchar[{\@Ewith}{\@Ewithout}}
\def\@Ewith[#1]#2{\mathbb{E}_{#1}\left[#2\right]}
\def\@Ewithout#1{\mathbb{E}\left[#1\right]}
\makeatother

\newcommand{\braces}[1]{\left\{#1\right\}}
\newcommand{\brackets}[1]{\left[#1\right]}
\newcommand{\parentheses}[1]{\left(#1\right)}
\DeclareMathOperator{\absWithoutParenth}{abs}
\newcommand{\abs}[1]{\absWithoutParenth\left(#1\right)}

\newcommand{\dist}{\sim}
\newcommand{\GammaDist}{\mathsf{Gamma}}
\newcommand{\NormalDist}{\mathsf{Normal}}
\newcommand{\UniformDist}{\mathsf{Uniform}}
\newcommand{\ie}{i.e.}
\newcommand{\eg}{e.g.}
\DeclareMathOperator{\var}{var}
\DeclareMathOperator{\trace}{tr}
\DeclareMathOperator{\softmax}{softmax}
\DeclareMathOperator{\argmax}{argmax}
\DeclareMathOperator{\argmin}{argmin}

\date{}

\begin{document}

\title[Unifying Summary Statistic Selection for Approximate Bayesian Computation]{Unifying Summary Statistic Selection for Approximate Bayesian Computation}

\author*[1]{\fnm{Till} \sur{Hoffmann}
}\email{thoffmann@hsph.harvard.edu}
\author[1]{\fnm{Jukka-Pekka} \sur{Onnela}
}\email{onnela@hsph.harvard.edu}

\affil[1]{\orgdiv{Department of Biostatistics}, \orgname{Harvard T.H. Chan School of Public Health}, \orgaddress{\street{677 Huntington Ave}, \city{Boston}, \state{Massachusetts} \postcode{02115}, \country{U.S.A}}}

\abstract{Extracting low-dimensional summary statistics from large datasets is essential for efficient (likelihood-free) inference. We characterize three different classes of summaries and demonstrate their importance for correctly analyzing dimensionality reduction algorithms. We demonstrate that minimizing the expected posterior entropy (EPE) under the prior predictive distribution of the model provides a unifying principle that subsumes many existing methods; they are shown to be equivalent to, or special or limiting cases of, minimizing the EPE. We offer a unifying framework for obtaining informative summaries and propose a practical method using conditional density estimation to learn high-fidelity summaries automatically. We evaluate this approach on diverse problems, including a challenging benchmark model with a multi-modal posterior, a population genetics model, and a dynamic network model of growing trees. The results show that EPE-minimizing summaries can lead to posterior inference that is competitive with, and in some cases superior to, dedicated likelihood-based approaches, providing a powerful and general tool for practitioners.}


\maketitle

\section{Introduction}

Empowered by advances in both scientific understanding and computing, researchers are developing ever more sophisticated simulators. For example, simulated weak lensing maps capture how dark matter affects light propagating through the universe \citep{Merten2019,Fluri2021}, coalescent simulators predict the evolution of genetic material \citep{Nordborg2019Coalescent}, and synthetic networks shed light on political opinion formation \citep{Sobkowicz2012}, effective vaccination strategies \citep{Yang2019}, and interactions between proteins \citep{Grassmann2024ProteinInteractions}.

While simulators can generate data $y$ given parameters $\theta$, we are often interested in the inverse problem: Constraining parameters $\theta$ given data $y$. If the likelihood $\likelihood{y\mid\theta}$ is available, we can use Markov chain Monte Carlo samplers \citep{Carpenter2017} or variational inference \citep[Ch.~10]{Bishop2006} to investigate the posterior $\posterior{\theta\mid y}$. But inference is more challenging if the likelihood is intractable or costly to evaluate.

Approximate Bayesian computation (ABC) overcomes this challenge in three steps by comparing observed with simulated data \citep{Beaumont2019}: First, we draw many samples $\parentheses{\theta_i,z_i}$ from the prior predictive distribution which form the so-called reference table. Second, we evaluate the distance $d_i=d\parentheses{y,z_i}$ between observed data $y$ and the $i^\mathrm{th}$ simulated dataset $z_i$. Finally, we accept $\theta_i$ as a sample from the ABC posterior $\abcposterior{\theta\mid y}$ if the distance $d_i$ is smaller than a threshold $\epsilon$. The smaller $\epsilon$, the better the approximation. Intuitively, ABC samples parameters $\theta_i$ that generate data $z_i$ which ``look like'' the observed data $y$. Hereafter, $y$ and $z$ will denote observed and simulated data.

Unfortunately, ABC suffers from the curse of dimensionality. The larger the dimensionality of the data, the larger the number of simulations required to obtain a sample that satisfies $d_i<\epsilon$. Compressing the data to lower-dimensional summary statistics $t=t(y)$ (or summaries in short) can overcome the curse of dimensionality but leaves us with the question: How do we choose the compression function $t(y)$?

\begin{figure}
    \centering
    \includegraphics{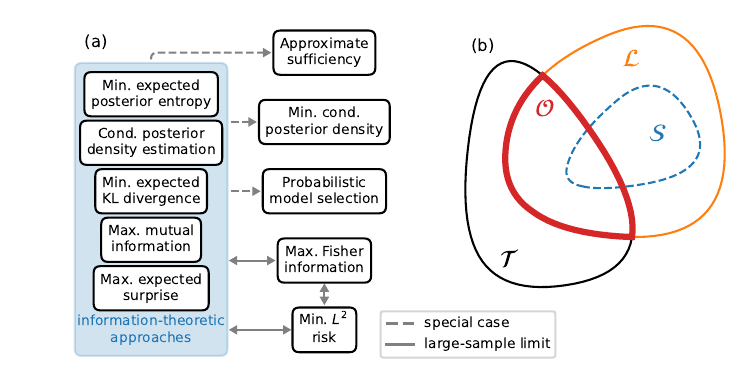}
    \caption{\emph{Different methods for compressing data to informative summaries are intimately related; distinguishing between classes of summaries is essential.} Panel~(a) illustrates that five information-theoretic approaches (ITAs) are equivalent. They implicitly minimize the same loss (\cref{sec:background,sec:epe}). Approximate sufficiency (\cref{sec:approximate-sufficiency}) seeks to achieve lossless compression, and minimizing the posterior entropy (\cref{sec:nunes}) is a special case of ITAs focused on only the observed data. Maximizing Fisher information (\cref{sec:fisher}) and minimizing $L^2$ Bayes risk (\cref{sec:bayes-risk}) are equivalent each other and ITAs in the large-sample limit. Probabilistic model selection (\cref{sec:model-selection}) maps onto ITAs if we treat model labels as parameters. A dashed arrow from one method to another indicates that the latter is a specialization of the former. Solid arrows indicate correspondence in the large-sample limit. Panel~(b) illustrates relationships between classes of summaries. Sufficient statistics $\mathcal{S}$ are a subset of lossless statistics $\mathcal{L}$ although the former only exist if the likelihood belongs to the exponential family.
    The intersection of lossless summaries $\mathcal{L}$ and the summaries $\mathcal{T}$ considered by the practitioner are optimal summaries $\mathcal{O}$. Optimal summaries are not necessarily lossless, \eg{} if $\mathcal{T}$ is restricted to parametric transformations.
    }
    \label{fig:relationships}
\end{figure}

A plethora of methods has been developed to address this question; some are summarized in panel~(a) of \cref{fig:relationships}. They include methods to select informative summaries from a pool of candidates \citep{Blum2010,Joyce2008,Nunes2010,Barnes2012,Blum2013} and parameterized transformations that can be optimized to learn summaries \citep{Aeschbacher2012,Fearnhead2012,Prangle2014,Jiang2017,Chan2018,Charnock2018,Chen2021MutualInfo,Radev2022BayesFlow}. Loss functionals quantifying how well the compressor preserves information have been motivated by minimizing the Bayes risk \citep{Fearnhead2012,Jiang2017}, model selection \citep{Prangle2014,Raynal2023,Merten2019}, and information theoretic arguments \citep{Nunes2010,Chen2021MutualInfo,Barnes2012,Charnock2018,Radev2022BayesFlow}.

We characterize three different classes summaries in \cref{sec:background}: sufficient, lossless, and optimal summaries. In \cref{sec:epe}, we argue that all information-theoretic approaches are equivalent. They implicitly minimize the same loss functional between the summary posterior $\posterior{\theta\mid t}$ given only $t$ and the true posterior $\posterior{\theta\mid y}$ given the entire dataset $y$. While these results are well established in information theory, they provide a unifying perspective of different summary extraction approaches. Minimizing the expected posterior entropy (EPE) should be the practitioner's choice because it is easier to evaluate than either the mutual information (MI) between model parameters and summaries or the Kullback-Leibler (KL) divergence between the posterior given the full data and posterior given only summaries. It also has strong connections with conditional posterior density estimation \citep{Papamakarios2016,Lueckmann2017}. But even methods developed to address different problems (such as parameter inference or model selection) in diverse fields (such as cosmology or population genetics), have strong ties to information-theoretic approaches. For example, in \cref{sec:related} we show that maximizing the determinant of the Fisher information \citep{Heavens2000,Charnock2018} and minimizing the $L^2$ Bayes risk \citep{Fearnhead2012,Jiang2017} are both equivalent to minimizing the EPE in the large-sample limit. Similarly, learning a probabilistic classifier for model selection \citep{Prangle2014} minimizes the EPE. In \cref{sec:experiments}, we discuss concrete steps for learning summaries by fitting conditional posterior density estimators to simulated data. To compare different methods, we devise a benchmark problem with simple likelihood but data that prove challenging for summary selection in \cref{sec:benchmark}. We also compare summary selection approaches on two applied examples: Inferring the mutation and recombination rates of a population genetics model (\cref{sec:population-genetics}) and the attachment kernel for a model of growing trees (\cref{sec:tree}).

\section{Background\label{sec:background}}

Given data $y$ we seek to infer parameters $\theta$ of a model using summaries $t=t(y)$ that retain as much information about the true posterior as possible. Summaries $t_\text{suff}$ with fixed and finite dimensions are \emph{Bayes sufficient} if $\posterior{\theta\mid t_\text{suff}}=\posterior{\theta\mid y}$ for all $y$ and any prior $\prior{\theta}$ \citep{Prangle2018}. But they only exist for exponential-family likelihoods \citep{Koopman1936}. We have to relax the concept of sufficiency, and we call statistics $t_\text{lossless}$ \emph{lossless} if
\begin{equation}
    \posterior{\theta\mid t_\text{lossless}(y)} = \posterior{\theta\mid y}\label{eq:lossless-defn}
\end{equation}
for all data $y$ of the same sample size, and a given prior $\prior{\theta}$. While lossless statistics always exist (\eg{} the identity map), they may not be useful in practice.
We say that the statistics $t_\text{opt}$ are \emph{optimal} if they minimize a non-negative loss functional that measures the discrepancy between the posterior given the full data and the posterior given only summaries. Specifically, we consider the loss functional
\begin{equation}
    \mathcal{L}_t=\int dz\,q\parentheses{z} \ell\braces{\posterior{\theta\mid z},\posterior{\theta\mid t(z)}},
    \label{eq:loss-functional}
\end{equation}
where $\ell$ is an instance-level loss functional that measures the discrepancy between true posterior $\posterior{\theta\mid z}$ and summary posterior $\posterior{\theta\mid t(z)}$ for a particular dataset $z$. Instance-level discrepancy measures $\ell$ include, for example, the KL divergence, Wasserstein distance, and total variation distance \citep{Cai2022}. As we discuss further in \cref{sec:maxinfo}, summaries that are informative for one dataset may be uninformative for another. The weighting function $q$ encodes which parts of the data space we prioritize. The optimal summaries are
\begin{equation}
    t_\text{opt}=\argmin_{t\in\mathcal{T}} \mathcal{L}_t,\label{eq:optimal-defn}
\end{equation}
where $\mathcal{T}$ is the space of summaries under consideration.
Consequently, sufficient statistics are lossless, and lossless statistics are optimal, but the converse is not necessarily true. For example, $\mathcal{T}$ may be restricted to parametric transformations \citep{Fearnhead2012} or selecting at most $k$ summaries from a set of candidate statistics \citep{Raynal2023}. The relationship between different classes of summaries is illustrated in panel~(b) of \cref{fig:relationships}.

The choice of summary statistic $t$ imposes a fundamental limit on the fidelity of the resulting posterior approximation irrespective of the ABC tolerance $\epsilon$. In the limit $\epsilon\rightarrow 0$, the distribution of accepted samples converges to the summary posterior $\posterior{\theta\mid t(y)}$. This distribution represents the best possible posterior approximation achievable with a given set of summaries. Consequently, even an ideal ABC procedure cannot recover information about the parameters that is lost during the initial data compression step. Minimizing the loss functional in \cref{eq:optimal-defn} improves this asymptotic target, ensuring that the best-case outcome of the inference is a high-fidelity approximation of the true posterior $\posterior{\theta\mid y}$.

Despite the pursuit of the holy grail of \emph{sufficient} statistics, we typically have to settle for the weakest concept of \emph{optimal} statistics. Even the most sophisticated method cannot extract sufficient statistics if the likelihood does not belong to the exponential family \citep{Koopman1936}. Similarly, unless the family of summaries $\mathcal{T}$ is rich enough, lossless compression is not achievable. Further, even if $\mathcal{T}$ is rich enough, one cannot in general verify that \cref{eq:lossless-defn} holds for all $\theta$ and $y$ given a finite computational budget.

While models with exponential-family likelihoods are theoretically appealing, they may not be sufficiently expressive or intuitive to address real-world problems. Domain knowledge can aid in the development of models that capture salient features of the data, including protein interaction networks \citep{Grassmann2024ProteinInteractions}, cosmology \citep{Charnock2018}, and population-genetics \citep{Nordborg2019Coalescent}. But these models often do not have sufficient statistics or even tractable likelihoods, and we need to resort to possibly lossy compression and likelihood-free inference.

\section{Minimizing the expected posterior entropy\label{sec:epe}}

A natural loss functional to minimize is the expected KL divergence from the true posterior $\posterior{\theta\mid z}$ to the summary posterior $\posterior{\theta\mid t(z)}$. Similar to the evaluation of the Fisher information \citep[Ch.~6]{Bishop2006}, the expectation is taken with respect to the prior predictive distribution $p(z)$ of the model, \ie $q(z)=p(z)$. This ensures that the summaries are informative for data that are plausible under the model. We propose choosing summaries that minimize the expected posterior entropy (EPE). This approach is equivalent to minimizing the expected KL divergence, conceptually simple, computationally tractable, and has a strong connection with recent inference techniques based on conditional density estimation \citep{Papamakarios2016,Lueckmann2017,Radev2022BayesFlow}.

The posterior entropy given summaries $t(z)$ for a fiducial dataset $z$ is
\begin{equation}
    \entropy\braces{\posterior{\theta\mid t(z)}}=-\int \diff{\theta} \posterior{\theta\mid t(z)}\log\posterior{\theta\mid t(z)}.\label{eq:posterior-entropy}
\end{equation}
Here, a fiducial dataset refers to a dataset generated based on known parameters.
Taking the expectation with respect to the data under the model yields the EPE
\[
    \epe{}\equiv\E[z\dist\proba{z}]{\entropy\braces{\posterior{\theta\mid t(z)}}}
    =-\int\diff{z}\diff{\theta}\proba{z}\posterior{\theta\mid t(z)}\log\posterior{\theta\mid t(z)},
\]
where $\proba{z}=\int \diff{\theta}\,\likelihood{z\mid\theta}\prior{\theta}$ is the marginal likelihood, and $\E[z\dist\proba{z}]{\cdot}$ denotes the expectation with respect to $z$ under the distribution $\proba{z}$. Changing variables of integration from data $z$ to summaries $t$ leaves us with the simple expression
\[
    \epe{}=-\int\diff{t}\diff{\theta} \proba{t,\theta}\log\posterior{\theta\mid t},
\]
where the Jacobian has been absorbed by the joint density $\proba{t,\theta}$. With a slight abuse of notation, we use $\proba{\cdot}$ for both the marginal likelihood and joint distribution where the distinction is unambiguous. Given a posterior density estimator $\hat f\parentheses{\theta\mid t}$ that seeks to approximate the summary posterior, we can construct a Monte Carlo estimate of the EPE
\begin{equation}
    \hat \epe{}=-m^{-1}\sum_{i=1}^m\log \hat f\parentheses{\theta_i \mid t(z_i)},\label{eq:epe-mc-estimate}
\end{equation}
where $\theta_i$ and $z_i$ are joint samples from $\proba{\theta,z}$, and $m$ is the number of samples. This estimate is the widely used loss function for learning the posterior from simulated data \citep{Papamakarios2016,Lueckmann2017,Radev2022BayesFlow}, where $m$ is the size of the mini-batch, \ie{} a subset of the data used to train the model.

We consider three well-established connections to other information-theoretic approaches \citep[Ch.~1]{Bishop2006} although with a specific focus on the selection of summaries for ABC. First, we evaluate the difference between the prior entropy and EPE
\begin{equation}
    \entropy\braces{\prior{\theta}} - \epe{} = \int \diff{t}\proba{t}\int\diff{\theta} \posterior{\theta\mid t} \log\parentheses{\frac{\posterior{\theta\mid t}}{\prior{\theta}}},\label{eq:prior-posterior-entropy-difference}
\end{equation}
where we have been able to combine the two integrals because
\[
    \int\diff{\theta}\prior{\theta}\log\prior{\theta}=\int\diff{t}\diff{\theta}\proba{t,\theta}\log\prior{\theta}
\]
by the law of total probability. The inner integral of \cref{eq:prior-posterior-entropy-difference} is the KL divergence from the prior to the posterior $\kldiv{\posterior{\theta\mid t}}{\prior{\theta}}$, sometimes called \emph{surprise} because it measures the degree to which an observer updates their belief in light of new data \citep{Itti2009}. Minimizing the EPE thus maximizes our expected surprise from observing the summaries because the prior entropy does not depend on the choice of summaries.

Second, we note that $\posterior{\theta\mid t}=\proba{t,\theta} / \prior{t}$ and \cref{eq:prior-posterior-entropy-difference} simplifies to the MI between the summaries $t$ and parameters $\theta$
\begin{equation}
    I\braces{\theta, t} = \int \diff{t}\diff{\theta} \proba{t,\theta} \log\parentheses{\frac{\proba{\theta, t}}{\prior{\theta}\proba{t}}}.\label{eq:mutual-info-defn}
\end{equation}
As the MI is non-negative, the EPE is not larger than the prior entropy, \ie{} we reduce uncertainty on average. Minimizing the EPE is equivalent to maximizing the MI which has been proposed in the context of subset selection \citep{Barnes2012} and neural summaries \citep{Chen2021MutualInfo}. However, estimating MI is difficult in high dimensions \citep{Jeffrey2020NeuralSummaries}, making the approach computationally challenging.

Third, we consider the difference between the EPE given only summaries $t$ and the EPE given a full fiducial dataset $z$
\[
    \epe{}-\E[z\dist\proba{z}]{\entropy\braces{\posterior{\theta\mid z}}}
    =\int \diff{z} \proba{z} \int \diff{\theta} \posterior{\theta\mid z} \log \parentheses{\frac{\posterior{\theta\mid z}}{\posterior{\theta\mid t}}},
\]
and we can identify the inner integral as the KL divergence from the summary posterior $\posterior{\theta\mid t}$ to the true posterior $\posterior{\theta \mid z}$ (see App.~\ref{app:epe-kl-relationship}). The difference of expected entropies is thus equal to the expected KL divergence between the posteriors
\[
    \epe{} - \E[z\dist\proba{z}]{\entropy\braces{\posterior{\theta\mid z}}}=\E[z\dist\proba{z}]{\kldiv{\posterior{\theta \mid z}}{\posterior{\theta \mid t}}},
\]
which \citet{Chan2018} used to infer recombination hotspots in population genetics and \citet{Radev2022BayesFlow} targeted for amortized Bayesian inference. Minimizing the EPE is equivalent to minimizing the expected KL divergence because the true posterior entropy given the complete dataset does not depend on the summaries. The KL divergence is non-negative which allows us to draw two conclusions. First, the EPE given only summaries $t\parentheses{z}$ is greater than or equal to the EPE given the full dataset $z$, \ie{} we lose information in expectation by conditioning on the summaries $t$ instead of the data $y$ unless the summaries are lossless. Second, minimizing the EPE implies that the loss functional in \cref{eq:optimal-defn} is the expected KL divergence. Similar to the MI, evaluating the expected KL divergence is challenging because neither the true posterior $\posterior{\theta\mid z}$ nor the summary posterior $\posterior{\theta\mid t\left(z\right)}$ are known in practice.

To summarize, minimizing the EPE, maximizing the MI between parameters $\theta$ and summaries $t$, maximizing the expected surprise, and minimizing the expected KL divergence between $\posterior{\theta\mid z}$ and $\posterior{\theta\mid t\left(z\right)}$ are equivalent, as illustrated in panel~(a) of \cref{fig:relationships}. But minimizing the EPE is preferable because it can be estimated using \cref{eq:epe-mc-estimate} for functional approximations of the posterior and nearest-neighbor entropy estimators for posterior samples \citep{Singh2003}.

\section{Related work and connections with expected posterior entropy\label{sec:related}}

\subsection{Approximate sufficiency\label{sec:approximate-sufficiency}}

\citet{Joyce2008} cast the task of selecting summaries as a sequence of hypothesis tests to select a subset of candidate summaries. Specifically, they considered
\[
    \log R_k\parentheses{\theta}=\log \abcposterior{\theta\mid t_{k},\ldots,t_1} - \log\abcposterior{\theta\mid t_{k-1},\ldots,t_1},
\]
where $\abcposterior{\theta\mid t_{k-1},\ldots,t_1}$ is the ABC posterior given $k-1$ summaries already selected and $\abcposterior{\theta\mid t_{k},\ldots,t_1}$ is the posterior resulting from including an additional statistic $t_k$. Intuitively, if the error score $\delta_k=\max_\theta \abs{\log R_k\parentheses{\theta}}$ is zero, \ie{} the two posteriors are identical, the $k^\mathrm{th}$ statistic does not capture additional information and can be ignored. If $\delta_k$ differs significantly from zero, we reject the null hypothesis that $\abcposterior{\theta\mid t_{k},\ldots,t_1}$ and $\abcposterior{\theta\mid t_{k-1},\ldots,t_1}$ are the same distribution and include $t_k$. They consider a set of $k-1$ summaries to be ``approximately sufficient'' if $\delta_k$ does not significantly differ from zero for any additional summary statistic.

This iterative process cannot minimize a loss functional of the form of \cref{eq:loss-functional} globally. Yet it approximately minimizes a loss functional that assigns all weight to the observed data $y$ and uses the maximum log density ratio to distinguish between true and summary posteriors as the instance-level loss functional, i.e.,
\begin{align*}
q(z)&=\delta\parentheses{z - y}\\
\ell &= \max_\theta\abs{\log\posterior{\theta\mid z}-\log\posterior{\theta\mid t(z)}},
\end{align*}
where $\delta$ denotes the Dirac delta function.

Importantly, the error score $\delta_k=\max_\theta \abs{\log R_k}$ assigns equal importance to all subsets of the parameter space, even regions we know to be irrelevant. For example, suppose that the posterior given the currently selected $k-1$ summaries is normal with variance $\sigma_{k-1}^2$, and the posterior after adding the $k^\mathrm{th}$ summary is identical except for a different variance $\sigma_{k}^2$. Even if $\sigma_k$ and $\sigma_{k-1}$ differ by an infinitesimal amount, $\delta_k$ is unbounded because
\[
    \delta_k = \frac{1}{2}\max_\theta\abs{\log\parentheses{\frac{\sigma_k^2}{\sigma_{k-1}^2}}+\parentheses{\frac{\sigma_{k-1}^2-\sigma_k^2}{\sigma_k^2\sigma_{k-1}^2}}\theta^2}=\infty.
\]
The error score is dominated by regions of the parameter space that have virtually no posterior mass. The expected value $\E[\theta\dist\posterior{\theta\mid t_k,\ldots,t_1}]{\log R_k\parentheses{\theta}}$ instead weights discrepancies between the two distributions by the posterior mass. This quantity is in fact the KL divergence considered by \citet{Barnes2012} (see \cref{sec:maxinfo} for details).

The notion of ``approximate'' sufficiency is necessarily a statement about limited computational resources: If we had unlimited resources, only candidate statistics that are uninformative or redundant would be excluded. This observation applies to any subset selection algorithm, such as minimizing posterior entropy \citep{Nunes2010} in \cref{sec:nunes}, regression-based subset selection methods \citep{Blum2010,Blum2013} in \cref{sec:bayes-risk}, or maximizing MI \citep{Barnes2012} in \cref{sec:maxinfo}.

\subsection{Minimizing the conditional posterior entropy\label{sec:nunes}}

\citet{Nunes2010} proposed choosing a subset of summaries $t$ by minimizing the conditional posterior entropy (CPE) $\entropy\braces{\posterior{\theta\mid t\parentheses{y}}}$ given data $y$. They ran rejection ABC for different subsets of summaries and evaluated the CPE using a nearest-neighbor estimator \citep{Singh2003}. The proposal is appealing because low-entropy posteriors give precise parameter estimates.

However, it implicitly assumes that the data we \emph{have} observed are the only data that could \emph{ever} be observed, similar to the non-parametric bootstrap. More formally, the weighting is $q(z)=\delta\parentheses{z-y}$ as in \cref{sec:approximate-sufficiency}, and the instance-level loss functional is the entropy of the summary posterior, \ie $\ell=\entropy{\posterior{\theta\mid t(z)}}$. When the maximum likelihood estimate of the parameters lies in the tail of the prior distribution, the CPE $\entropy\braces{\posterior{\theta\mid y}}$ can be larger than the prior entropy $\entropy\braces{\prior{\theta}}$ because the true posterior is a ``compromise'' between prior and likelihood \citep{Blum2013}.

\begin{figure}
    \centering
    \includegraphics{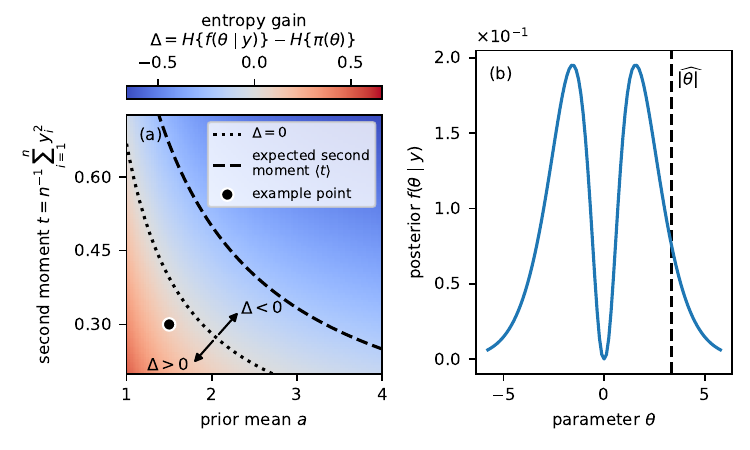}
    \caption{\emph{Extracting summaries can be non-trivial even for toy models.} Panel~(a) shows the difference between posterior and prior entropy for a model with zero-mean normal likelihood and conjugate gamma prior for the precision $\theta$ (inverse variance). For a subset of the prior and data space, minimizing the posterior entropy discards the second moment $t$, a sufficient statistic. Panel~(b) shows the bimodal posterior for the example point in~(a) that arises when the precision of the likelihood is $\abs{\theta}$ (see \cref{sec:bayes-risk}). The posterior mean is zero and not informative of the parameter. The vertical dashed line represents the maximum likelihood estimate $\widehat{\abs{\theta}}$ of the precision $\abs{\theta}$.}
    \label{fig:tough-nuts}
\end{figure}

We consider a simple example with closed form posterior because it illustrates important concepts and challenges associated with learning summaries. Suppose we draw $n=4$ samples $y$ from a zero-mean normal distribution with unknown precision (inverse variance) $\theta$. We use a gamma prior for $\theta$ because it is the conjugate prior for a normal likelihood with known mean. The distribution is parameterized by a shape parameter $a$ and rate parameter $b$. We use $b=1$ such that the prior mean is $a$. More formally,
\begin{align*}
    \theta\mid a, b&\dist\GammaDist\parentheses{a, b}\\
    y_i\mid\theta&\dist\NormalDist\parentheses{0, \theta^{-1}},
\end{align*}
where $i\in\braces{1,\ldots,n}$. The closed-form posterior is
\begin{equation}
    \theta\mid y, a, b \dist\GammaDist\parentheses{a + \frac{n}{2}, b + \frac{n t}{2}},
    \label{eq:precision-posterior}
\end{equation}
where $t=n^{-1}\sum_{i=1}^n y_i^2$ is the second moment, a sufficient statistic. For example, if $a=1.5$ and $t=0.3$, the prior entropy is $1.36$ and the CPE is $1.47$. Minimizing the CPE would discard the sufficient statistic $t$ such that the posterior is equal to the prior: We have not learned anything from the data. Panel~(a) of \cref{fig:tough-nuts} shows the entropy gain $\Delta=\entropy\braces{\posterior{\theta\mid y}}-\entropy\braces{\prior{\theta}}$ in light of the data for different priors and sample variances. Indeed, generating $10^5$ samples from the prior predictive distribution with $a=1.5$, we find that $30\%$ of samples lead to a CPE increase. Interestingly, this situation is more likely to arise when the ``surprise'' \citep{Itti2009} is large, and we should substantially update our beliefs in light of the data. In contrast, the EPE $\epe=0.87$ given $t$ is smaller than the prior entropy, and minimizing it would select $t$ as a useful summary. Monte Carlo standard errors of the EPE and proportion of entropy increases are smaller than the reported significant digits.

The instance-level loss functional, the entropy of the summary posterior, is not a discrepancy measure between the true and summary posteriors, and \citet{Nunes2010} also considered a two-stage method: First they used the above approach to select candidate summaries and identify simulated datasets close to the observed data. Second, they drew posterior samples for each identified dataset and evaluated the root mean integrated squared error (RMISE) of posterior samples for each subset of summaries. This is possible because the parameters of simulated datasets are known. The summaries with the lowest RMISE were then selected. We do not consider this two-stage approach further here because of its computational burden and because posterior mean estimation methods optimize a similar objective, as discussed in \cref{sec:bayes-risk}.

\subsection{Maximizing the Fisher information\label{sec:fisher}}

Even when the likelihood is tractable, compressing the data $y$ to summaries $t$ has computational benefits. \citet{Heavens2000} developed an optimal linear compression scheme for Gaussian likelihoods in the sense that the Fisher information is preserved. Information-maximizing neural networks \citep{Charnock2018} seek to maximize the determinant of the Fisher information matrix when linear compression is not sufficient, and methods to maximize the Fisher information for non-Gaussian likelihoods have recently been developed \citep{Alsing2018,Fluri2021}. Fisher information methods are fundamentally likelihood-based and do not fit into the loss functional framework of \cref{eq:loss-functional}. However, we can establish a connection to minimizing the EPE in the large-sample limit.

We consider the large-sample limit $n\rightarrow\infty$ of $n$ i.i.d. observations $z=\parentheses{z_1,\ldots,z_n}$ and summaries of the form $t\parentheses{z}=n^{-1}\sum_{i=1}^n h\parentheses{z_i}$ where $h$ is a potentially nonlinear function. This restriction preserves the i.i.d.\ structure required for the Bernstein--von Mises theorem and is consistent with the observation that summaries often have well-behaved likelihoods when they are means of i.i.d.\ data \citep{Alsing2018}. According to the Bernstein--von Mises theorem, the posterior approaches a multivariate normal distribution under certain regularity conditions \citep{Vaart1998}. Specifically,
\[
    \theta\mid t\dist\NormalDist\parentheses{\theta_0,F^{-1}\parentheses{\theta_0}},
\]
where $\theta_0$ is the true parameter that generated the summaries $t$, and
\begin{equation}
F_{ij}\parentheses{\theta_0}=\E[z\dist\proba{z}]{\parentheses{\frac{\partial}{\partial\theta_i} \log\likelihood{t(z)\mid\theta}}\parentheses{\frac{\partial}{\partial\theta_j}\log\likelihood{t(z)\mid\theta}}}_{\theta=\theta_0}
\label{eq:fisher-information}
\end{equation}
is the Fisher information of the summaries evaluated at $\theta_0$ \citep[Ch.~6]{Bishop2006}. The limiting entropy of the posterior can thus be readily evaluated and is
\[
    \lim_{n\rightarrow\infty}\entropy\braces{\posterior{\theta\mid t}}=-\frac{1}{2}\log\det F\parentheses{\theta_0} + \text{constant},
\]
where $\det F$ denotes the determinant of $F$. We take the expectation with respect to the prior $\pi$ to obtain the EPE
\[
    \lim_{n\rightarrow\infty}\epe = -\frac{1}{2}\int \diff{\theta_0} \prior{\theta_0}\log\det F\parentheses{\theta_0} + \text{constant}.
\]
We do not need to take an expectation over summaries $t\mid\theta_0$ because the Fisher information in \cref{eq:fisher-information} does not depend on the realization $t$. Maximizing the expected log determinant of the Fisher information matrix is thus equivalent to minimizing the EPE in the large-sample limit. This observation agrees with our intuition that the effect of the prior on the posterior decreases as the sample size increases.

We argue that minimizing the EPE is more appealing than maximizing the Fisher information for three reasons. First, it can incorporate prior information in the small-$n$ regime to yield the most faithful posterior approximation. Second, it does not require the choice of a fiducial value of $\theta$ at which to evaluate the Fisher information. Finally, when the likelihood is not available, we need to  approximate it to evaluate the Fisher information. For example, \citet{Charnock2018} assume that the likelihood of the learned summaries can be approximated by a Gaussian, and \citet{Alsing2018} argue that candidate summaries often have a Gaussian likelihood if they are the mean of i.i.d.\ data.

\subsection{Minimizing the Bayes risk\label{sec:bayes-risk}}

\citet{Fearnhead2012} proposed the posterior mean of the parameters as summaries. Of course, the posterior mean is not known, but we can estimate it by minimizing the quadratic loss
\begin{equation}
    \ell=\E[z,\theta\dist \proba{z,\theta}]{\parentheses{\theta-t_\beta(z)}^\intercal A\parentheses{\theta-t_\beta(z)}}\label{eq:quadratic-loss}
\end{equation}
where $t_\beta(z)$ is a predictor of $\theta$ parameterized by $\beta$, $A$ is a positive-definite matrix, and $^\intercal$~denotes the transpose. The approach fits into the loss functional framework of \cref{eq:loss-functional} with $q(z)=\proba{z}$ (the prior predictive distribution) and instance-level loss functional
\[
\ell=\int\diff{z}\posterior{\theta\mid z}\parentheses{\theta-t_\beta(z)}^\intercal A\parentheses{\theta-t_\beta(z)},
\]
where $t$ is constrained to be the posterior mean. \citet{Fearnhead2012} considered linear predictors, but neural networks \citep{Jiang2017} and boosted regression \citep{Aeschbacher2012} have also been proposed. In practice, the parameters $\beta$ are learned by minimizing a Monte Carlo estimate of \cref{eq:quadratic-loss} akin to \cref{eq:epe-mc-estimate}. Using the estimated posterior mean $t_\beta\parentheses{\cdot}$ as summaries implicitly chooses as many summaries as there are parameters.

Considering again the large-sample limit, the quadratic loss becomes (adapted from Theorem~3 of \citet{Fearnhead2012})
\[
    \ell=\trace\brackets{A\int\diff{\theta}\prior{\theta}F^{-1}\parentheses{\theta}},
\]
where $\trace$ denotes the matrix trace. Consequently, minimizing the quadratic loss in \cref{eq:quadratic-loss} is intimately related to maximizing the determinant of the Fisher information because both $A$ and $F$ are positive-definite. However, the details depend on the form of $A$.

The above argument crucially depends on the assumptions of the Berstein--von Mises theorem holding. In particular, the model needs to be identifiable such that different values of the parameters $\theta$ are distinguishable in the $n\rightarrow\infty$ limit \citep{Vaart1998}. We consider a variant of the toy model presented in \cref{sec:nunes} that is not identifiable and discuss the impact on learning summaries. In particular, we use the absolute value $\abs{\theta}$ of a parameter $\theta$ as the precision such that the conditional distributions are
\begin{align*}
    \abs{\theta}\mid a,b&\dist \GammaDist\parentheses{a,b}\\
    y_i&\dist\NormalDist\parentheses{0,\abs{\theta}^{-1}}.
\end{align*}
The real-valued $\theta$ is distributed as a mixture of a gamma distribution and its reflection about the origin under the prior. The closed-form posterior is
\[
    \abs{\theta}\mid y,a,b \dist \GammaDist\parentheses{a+\frac{n}{2}, b+\frac{nt}{2}},
\]
where $t$ is the second moment of $y$ as in \cref{eq:precision-posterior} and a sufficient statistic. The posterior is bimodal and symmetric under reflection, as shown in panel~(b) of \cref{fig:tough-nuts}. The posterior mean is zero, and it is not possible to extract information by minimizing \cref{eq:quadratic-loss}.

This example may seem contrived, but multimodal posteriors that render the posterior mean uninformative are not uncommon. For example, mixture models are invariant under label permutation \citep{Stephens2000}, and latent-space models of networks \citep{Hoff2002} as well as latent factor models for Bayesian PCA \citep{Nirwan2019} are invariant under rotations. The limitation of the Bayes risk approach arises because the instance-level loss functional measures concentration around a point rather than comparing full posterior distributions. Using information theoretic approaches ensures we stay focused on the task at hand: Approximating the true posterior.

The relationship between parameters and data can be complex, and regression approaches, especially linear regression, may not be able to capture the relationship globally. Local relationships in regions of high posterior mass can be learned using pilot runs \citep{Fearnhead2012} or weighting samples \citep{Blum2010}. Local regression methods have also been adapted for subset selection: A model is fit to predict parameters from candidate summaries, and a candidate is selected if it increases a metric such as the Bayesian evidence \citep{Blum2010}, Akaike information criterion, or Bayesian information criterion \citep{Blum2013}.

\subsection{Maximizing the mutual information\label{sec:maxinfo}}

\citet{Barnes2012} proposed choosing summaries from a pool of candidates that maximize the MI $I\braces{\theta,t}$ between parameters $\theta$ and the statistics $t$. Assuming that the candidate set includes sufficient statistics $t_\text{suff}$ such that
\begin{equation}
    \posterior{\theta\mid t_\text{suff}}=\posterior{\theta\mid y}
    \label{eq:sufficient-statistics-defn}
\end{equation}
for all possible $y$, they constructed a set of summaries sequentially. At the $k^\mathrm{th}$ step, they included the summary that maximizes the surprise given the $k-1$ statistics that have already been selected. The approach is similar to the approximate sufficiency method reviewed in \cref{sec:approximate-sufficiency}, but candidates are prioritized by their surprise at each stage. Together, the steps select the summaries that maximize the surprise $\kldiv{\posterior{\theta\mid t}}{\prior\theta}$ for the observed data. Like \citet{Joyce2008} and \citet{Nunes2010}, this approach considers only the observed dataset with $q(z)=\delta(z-y)$ in the loss functional framework of \cref{eq:loss-functional}. Consequently, it maximizes the conditional surprise $\ell=\kldiv{\posterior{\theta\mid t(y)}}{\prior\theta}$ rather than the MI, which is the expected surprise under the prior predictive distribution.

However, recall from \cref{eq:prior-posterior-entropy-difference,eq:mutual-info-defn} that the MI is equal to the \emph{expected} surprise under the generative model. In general, maximizing the surprise for a particular observed dataset is thus not equivalent to maximizing the MI. The approach may select different summaries if the candidate set does not include sufficient statistics.

Similarly, \citet{Chen2021MutualInfo} sought to maximize the MI using a neural network. They suggested that ``$t(z)$ is a sufficient statistic for $\likelihood{z\mid\theta}$ if and only if''~(p.~2) it maximizes the MI and ``that the sufficiency of the learned statistics is insensitive to the choice of $\prior{\theta}$''~(p.~4) such that ``[their approach] is globally sufficient for all $\theta$''~(p.~6)\footnote{We have adapted notation in quotations for consistency with this manuscript.}. As we shall illustrate with a toy model, these propositions do not hold in general because of the difference between sufficient and optimal statistics discussed in \cref{sec:background} (see App.~\ref{app:chen} for details).

\begin{figure}
    \centering
    \includegraphics{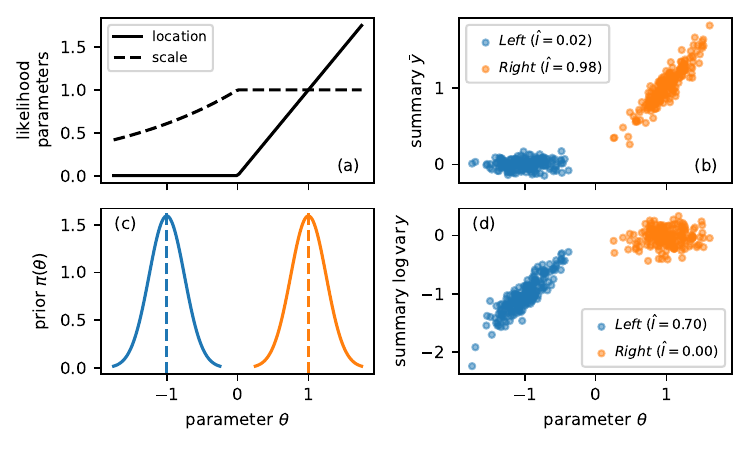}
    \caption{\emph{Optimal summaries depend on the prior.} Panel~(a) shows the parameters of a piecewise likelihood with qualitatively different behaviour on either side of the transition at $\theta=0$. Panel~(c) shows two priors with the bulk of their mass on either side of the transition. Panels~(b) and~(d) show the relationship between the parameter and the sample mean $\bar y$ and log variance $\log\var y$, respectively, as a scatter plot. Mutual information estimates highlight that the optimal choice of summary depends on the prior: The $\bar y$ and $\log\var y$ summaries are informative for the priors centred at $+1$ and $-1$, respectively.}
    \label{fig:piecewise}
\end{figure}

Consider the piecewise likelihood
\begin{equation}
    y_i\mid\theta \dist \begin{cases}
    \NormalDist\parentheses{0, \exp\theta}&\text{if }\theta < 0\\
    \NormalDist\parentheses{\theta,1}&\text{if }\theta\geq 0
    \end{cases}\label{eq:piecewise-likelihood}
\end{equation}
which is continuous at the transition, as illustrated in panel~(a) of \cref{fig:piecewise}. We consider two different normal priors with common standard deviation of $0.25$ centred at $\pm 1$, as shown in panel~(c). For the purpose of this example, we may choose one summary from the candidate set comprising the sample mean $\bar y$ and the natural logarithm of the sample variance $\log\var y$, \ie{} we restrict the space of compression functions $\mathcal T$\footnote{We restrict $\mathcal{T}$ to simplify the example. Together, the two summaries are sufficient. A continuous mixture of the ``left'' and ``right'' part of the likelihood with logistic mixture weight would yield similar results but does not have sufficient statistics.}. Intuitively, the latter is informative for the ``left'' region of the parameter space and the former for the ``right''. This intuition is confirmed by simulation: We consider $m=10^5$ independent samples from each prior and draw $n=100$ observations from the likelihood in \cref{eq:piecewise-likelihood}. The relationship between the parameter $\theta$ and sample mean as well as log sample variance are shown in panels~(b) and~(d), respectively. For quantitative comparison, we also estimate the MI for all pairs of priors and summaries using a nearest-neighbor entropy estimator \citep{Singh2003}. On the one hand, the log sample variance ($\hat I=0.70$) is the optimal summary for the left prior because the sample mean provides little information ($\hat I=0.02$). On the other hand, the sample mean is highly informative for the right prior ($\hat I=0.98$) whereas the log sample variance is not informative ($\hat I=0.00$). As Bayesians, we cannot escape the prior, and the optimal summaries depend on it.

\subsection{Model selection\label{sec:model-selection}}

\citet{Prangle2014} used logistic regression to learn summaries that can discriminate between different models: The predicted class probabilities. Similarly, \citet{Merten2019} applied deep convolutional neural networks to weak lensing maps to learn features that can discriminate between nine different cosmological models, although not in the context of ABC. Such probabilistic approaches to model classification are equivalent to minimizing the EPE: Consider a one-hot encoding of the model index such that $\theta_j = 1$ if model $j$ generated the data and $\theta_j=0$ otherwise. The log summary posterior is thus
\begin{equation}
    \log\posterior{\theta\mid t}=\sum_{j=1}^r \theta_j \log\posterior{\theta_j=1\mid t},\label{eq:softmax-cross-entropy}
\end{equation}
where $r$ is the number of models under consideration, and $\posterior{\theta_j=1\mid t}$ is the posterior probability that the data were generated by model $j$. \Cref{eq:softmax-cross-entropy} is familiar as the negative cross-entropy loss widely used for multiclass classification in machine learning \citep[Ch.~4]{Bishop2006}. In other words, any machine learning classifier that minimizes the cross-entropy loss for model selection minimizes the EPE of the model labels.

\subsection{Conditional posterior density estimation\label{sec:conditional-posterior-density}}

As briefly discussed in \cref{sec:epe}, recent approaches to likelihood-free inference based on conditional density estimation minimize the EPE using the mini-batch estimator in \cref{eq:epe-mc-estimate} as a loss function \citep{Papamakarios2016,Lueckmann2017,Radev2022BayesFlow}. These methods are appealing because they can automatically compress large datasets although at the cost of having to choose an architecture for the density estimator which is an active area of research \citep{Papamakarios2021}. \citet{Chan2018} proposed exchangeable neural networks such that the output is invariant under permutations of i.i.d. data generated by the model. While neural density estimators can in principle learn such invariances, it is beneficial to encode symmetries in the architecture to improve efficiency and reduce the amount of training data required. In practice, conditional density estimators can have computational advantages over rejection ABC because they interpolate the posterior density in the parameter space, requiring fewer simulations \citep{Papamakarios2016}. However, they cannot offer the same asymptotic guarantees as ABC: As the tolerance parameter of the acceptance kernel is reduced, the sampling distribution converges to the summary posterior \citep{Beaumont2019}.

\subsection{Partial least squares regression}
\label{sec:further-methods}

\citet{Wegmann2009PartialLeastSquares} obtained summaries using partial least squares regression (PLSR), a latent variable model for supervised dimensionality reduction. The method projects data $z$ (or candidate summaries) to a latent space such that the embeddings are most predictive of the parameters $\theta$ as measured by the $L_2$ norm. Instead of the predictions of the model as in \cref{sec:bayes-risk}, the latent variables are used as summaries. The number of latent components is chosen using leave-one-out cross-validation based on the ability of the model to predict parameters. Similar to the subset selection methods discussed in \cref{sec:approximate-sufficiency,sec:nunes}, the number of components chosen by cross-validation is determined by computational constraints: For sufficiently large reference tables, the dimensionality of the candidate summaries is maintained provided each candidate summary encodes some information, however weak.

\section{Experiments\label{sec:experiments}}

\subsection{Evaluation criteria and model architecture for nonlinear methods}

We consider three experiments to compare methods for extracting summaries: A benchmark model with i.i.d. observations and tractable likelihood in \cref{sec:benchmark}, a population genetics model with data comprising candidate summaries in \cref{sec:population-genetics}, and a model of growing trees in \cref{sec:tree}. We first discuss the approach for consistently evaluating summary extraction methods and subsequently consider each experiment in depth.

For subset selection methods (such as minimizing the CPE in \cref{sec:nunes}) and simple projections (such as linear posterior mean estimation in \cref{sec:bayes-risk}), we evaluated candidate summaries that were supplied to each algorithm. For more flexible nonlinear posterior mean estimation, we developed experiment-specific neural compressors $t:\mathbb{D}\rightarrow \mathbb{R}^q$ to compress the raw data $z\in\mathbb{D}$ to $q$ low-dimensional summaries. The networks were trained by minimizing the quadratic loss in \cref{eq:quadratic-loss}.

\begin{figure}
    \centering
    \includegraphics{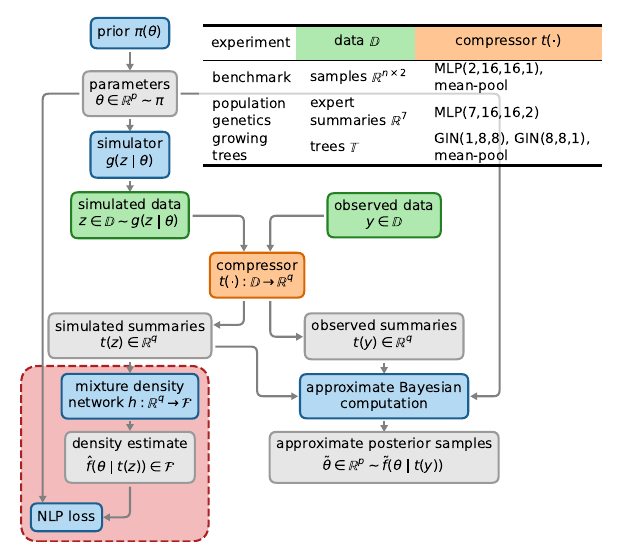}
    \caption{\emph{Mixture density networks with a bottleneck can learn informative summaries.} The stack left of the compressor $t$ illustrates the training data generation and MDN training procedure: $p$-dimensional parameters $\theta$ and synthetic data $z$ are drawn from the prior $\pi$ and simulator $g$, respectively. Synthetic data are compressed to summaries using a compressor $t$. The stack right of the compressor $t$ illustrates approximate Bayesian computation using learned summaries: The compressor evaluates summaries of observed data $y$, and parameter samples are accepted if corresponding simulated summaries $t\parentheses{z}$ are sufficiently close to observed summaries $t\parentheses{y}$. The red dashed box indicates components specific to training MDN compression: A mixture density network (MDN) $h$ estimates a posterior approximation $\hat f\parentheses{\theta\mid t(z)}$ given learned summaries $t(z)$. Here, $\mathcal{F}$ are the supported posteriors, \eg{} MDNs with certain component distributions. The network is trained by minimizing the negative log probability (NLP) loss. The table lists the type of data $\mathbb{D}$ and compressor architecture for each experiment (see \cref{sec:benchmark,sec:population-genetics,sec:tree} for details).}
    \label{fig:experiments}
\end{figure}

Summaries minimizing the EPE are appealing, but a concrete algorithm is required to make them useful in practice. We employed a conditional mixture density network (MDN) \citep{Papamakarios2016} with a bottleneck akin to an autoencoder \citep{Kramer1991Autoencoder}. The network comprises two parts: First, for fair comparison, we used the same neural compressor as for nonlinear posterior mean estimation. Consequently, the number of summaries $q$ is equal to the number of parameters $p$, although, in general, a different number of summaries $q>p$ could be chosen \citep{Chen2021MutualInfo}. Second, we extended the network with a conditional MDN to estimate the posterior density given only the summaries. The whole network comprising compressor and MDN was trained end-to-end by minimizing the Monte Carlo estimate of the EPE defined in \cref{eq:epe-mc-estimate}. After training, the bottleneck architecture ensures any information that may be useful for minimizing the EPE is captured by the output of the compressor; we dub this approach MDN compression. The inference pipeline for all methods is illustrated in \cref{fig:experiments}. For MDN compression, the compressor is trained by jointly optimizing a mixture density network on simulated data (shown in red) to minimize the expected posterior entropy. Once trained, summaries are extracted and used in ABC like other methods.
A similar approach was used by \citet{Jeffrey2020NeuralSummaries} for summaries fed to a likelihood estimation network. \citet{Radev2022BayesFlow} used a similar architecture of compression and density estimation networks, although using a normalizing flow for the latter. They used 128 summaries which is prohibitively large for ABC.

For each experiment, we generated a training, validation, and test set by sampling from the prior predictive distribution. Neural compressors were trained by mini-batch gradient descent using the Adam optimizer with default parameters and an initial learning rate of $10^{-2}$ \citep{Kingma2015}. The learning rate was decreased by an order of magnitude if the loss evaluated on the validation set did not decrease for ten consecutive epochs; training was stopped if it did not decrease for twenty consecutive epochs.

After extracting summaries for each example in the test set, we obtained samples from the approximate posterior $\tilde f\parentheses{\theta\mid t\parentheses{y}}$ in three steps: First, to ensure a common scale across summaries, we standardized them independently to have zero mean and unit variance based on the training set. Second, we evaluated the Euclidean distance $d_i$ between standardized summaries of each example $y$ and the $i^\text{th}$ element of the training set $z_i$. Third, we accepted a small fraction of the training set as posterior samples such that they had the smallest distance to each example \citep{Beaumont2019}, \ie{} the training set served as the reference table. The same training, validation and test sets were used for different methods for fair comparison. In addition to ABC, we drew samples from the prior as a baseline as well as directly from the MDNs trained to obtain EPE-minimizing summaries.

We used two metrics to evaluate approximate posterior samples. First, the root mean integrated squared error (RMISE)
\begin{equation}
    \text{RMISE} = \brackets{\frac{1}{s}\sum_{i=1}^s \left\Vert\tilde\theta_i -\theta\right\Vert^2}^{1/2},
    \label{eq:rmise}
\end{equation}
where $\tilde\theta_i$ denotes the $i^\text{th}$ sample from the ABC posterior and $s$ is the number of samples. This metric has been widely used in the ABC literature to evaluate summary extraction methods \citep{Joyce2008,Nunes2010,Fearnhead2012,Blum2013,Burr2013Calibration,Jiang2017}. It measures how concentrated ABC posterior samples are around the true parameter value $\theta$ \citep[Ch.~3]{Bishop2006}. The RMISE is a suitable metric for unimodal but not multimodal posteriors, as illustrated in panel~(b) of \cref{fig:tough-nuts}. Second, to address this shortcoming, we also evaluated the negative log probability (NLP) using kernel density estimation. Specifically,
\[
\text{NLP} = -\log\brackets{\frac{1}{s}\sum_{i=1}^s K_h\left(\tilde\theta_i-\theta\right)},
\]
where $K_h$ is a Gaussian kernel with bandwidth $h$ chosen by Scott's rule \citep{Scott2015MultivariateDensityEstimation}. For each experiment, metrics reported in \cref{fig:evaluation} and \cref{tbl:methods} in the appendix are averaged over the corresponding test set.

\subsection{Benchmark model\label{sec:benchmark}}

\begin{figure}
    \centering
    \includegraphics{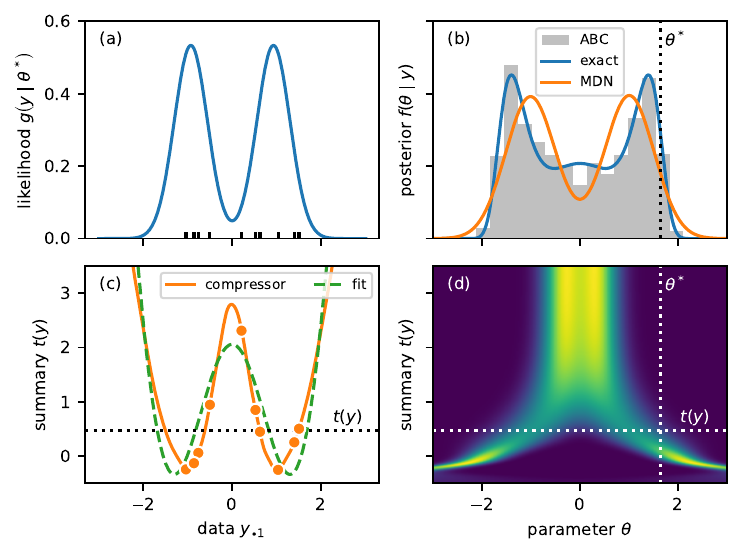}
    \caption{\emph{A conditional mixture density network (MDN) that minimizes the expected posterior entropy learns highly informative summaries.} Panel~(a) shows the likelihood for the true parameter $\theta^*\approx 1.6$ that generated the example dataset $y$ together with a rug plot for the $n=10$ observations $y_{\bullet 1}$. Panel~(b) shows the true posterior $\posterior{\theta\mid y}$ together with the learned posterior density estimator. While the two-component mixture is not flexible enough to approximate the true posterior well, it learns highly informative summaries: MDN-compressed ABC samples using these summaries are shown as a histogram. Panel~(c) shows the learned summary function $t: \mathbb{R}^{10 \times 2} \to \mathbb{R}$ which maps the full data matrix to a scalar; the plot shows $t(y)$ as a function of the first column values $y_{\bullet1}$ (the informative data, with the second column being uninformative noise). The dashed line shows how $t$ can be approximated using polynomial basis functions of the candidate summaries (the first three even moments). Panel~(d) illustrates the relationship between the posterior density estimator and the summary as a heat map; lighter colours indicate higher posterior density.}
    \label{fig:benchmark}
\end{figure}

We considered a benchmark model with multimodal true posterior set up to be challenging for extracting summaries. The model has a tractable likelihood that allowed us to compare the posterior given summaries with true posterior samples. In particular, we sampled a univariate parameter $\theta$ from the standard normal distribution and drew $n=10$ independent samples from the mixture distribution
\[
    z_{i1}\mid\theta \dist \frac{1}{2}\sum_{u\in\braces{-1,1}}\NormalDist\parentheses{u\times \tanh \theta, 1 - \tanh^2\theta},
\]
as illustrated in panel~(a) of \cref{fig:benchmark}. We also sampled a standard normal distractor $z_{i2}$ (uninformative noise) for each observation $i$ such that the full dataset $z=[z_{ij}]$ is a matrix with $n$ rows and $2$ columns. Learning or selecting summaries is non-trivial because all elements of $z$ have zero mean and unit variance under the generative model irrespective of the parameter $\theta$. The first moment is zero by symmetry; the second moment of each mixture component is $\E{z_{i1}^2}=\E{z_{i1}}^2+\var z_{i1}=\tanh^2\theta + 1 - \tanh^2\theta=1$ such that the mixture has unit variance. Sampling from the prior predictive distribution, we generated training, validation, and test sets of $10^6$, $10^4$, and $10^3$ independent realizations, respectively. The test set was used to evaluate and compare different methods. We employed the likelihood-based inference framework Stan \citep{Carpenter2017} to draw $1{,}000$ posterior samples for each example in the test set (see App.~\ref{app:benchmark-stan} for details). These samples formed the gold standard which we compared other methods to. \Cref{fig:benchmark} illustrates the learned summaries for a particular example dataset $y$ generated with true parameter $\theta^* \approx 1.6$.

For ABC using candidate summaries, the CPE minimization method \citep{Nunes2010}, PLS \citep{Wegmann2009PartialLeastSquares}, and linear posterior mean estimation \citep{Fearnhead2012}, we used the first three even moments of each column of $z$ as candidate summaries, giving rise to six statistics in total. Odd moments are not informative as the likelihood is symmetric, and we did not include them in our set of candidate summaries.

For the nonlinear posterior mean approximation \citep{Jiang2017}, we used a multilayer perceptron (MLP) that acts on each row of $z$ independently before compressing to a scalar summary. This architecture shares weights across all observations and is permutation invariant \citep{Chan2018}. The MLP has three fully connected layers (16, 16, and 1 hidden units) followed by mean-pooling across the $n$ observations; we used $\tanh$ activation functions throughout. The network was implemented in PyTorch \citep{Paszke2019} and trained as described in the preceding section with a mini-batch size of $512$.

Finally, we used a conditional MDN with $k=2$ Gaussian components to estimate the posterior and learn MDN-compressed summaries \citep{Bishop1994,Papamakarios2016}. To evaluate mixture logits $\eta(t)$, locations $\mu(t)$, and log-scales $\kappa(t)$ as a function of the summary $t$ we used independent two-layer MLPs (16 and $k$ hidden layers). The posterior density estimator is thus
\[
    \hat{f}\parentheses{\theta\mid t}=\sum_{j=1}^k \softmax_j \parentheses{\eta\parentheses{t}}\times\NormalDist\parentheses{\theta\mid\mu_j\parentheses{t},\exp\parentheses{2\kappa_j\parentheses{t}}},
\]
where $\softmax_j \parentheses{\eta} = \exp\eta_j / \sum_{l=1}^k\exp\eta_l$.

\begin{figure}
    \centering
    \includegraphics{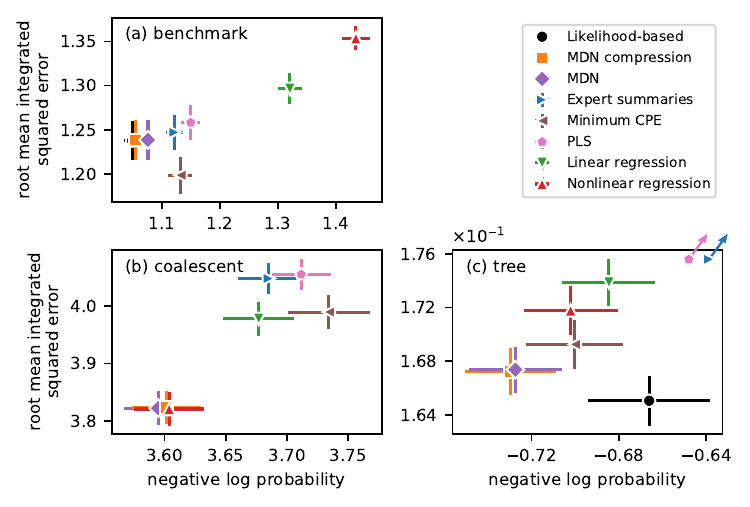}
    \caption{\emph{The quality of summaries significantly impacts the fidelity of posteriors.} Panels~(a), (b), and~(c) report the negative log probability loss and root mean integrated squared error of different methods for the benchmark, coalescent, and growing tree experiments, respectively. ``MDN'' refers to drawing samples directly from the learned mixture density network, ``likelihood-based'' refers to samples obtained with Stan for the benchmark problem and pseudo-marginal MCMC for the growing tree model, and all other methods use ABC after extracting summaries. Error bars are standard errors based on a test set of 1{,}000 i.i.d. samples for each experiment. Expert summaries and PLS perform poorly for the growing tree experiment and are indicated as off-the-chart by arrows. See \cref{sec:benchmark,sec:population-genetics,sec:tree} for details on the expert summaries for the benchmark, coalescent, and growing tree models, respectively.}
    \label{fig:evaluation}
\end{figure}

A comparison of the performance of different methods based on $1{,}000$ approximate posterior samples ($0.1\%$ of the reference table) is shown in panel~(a) of \cref{fig:evaluation} (see \cref{tbl:methods} in the appendix for a table of results). We report the RMISE for completeness, but it is a poor metric for multimodal posteriors. For example, a point mass at $\theta=0$ would have $\text{RMISE} = 1$---lower than any of the methods we considered. As expected, linear and nonlinear posterior mean estimators performed worst in terms of NLP because the posterior is bimodal. Because of its flexibility, the nonlinear estimator was able to accurately estimate the posterior mean $\E[\theta\dist\posterior{\theta\mid z}]{\theta}=0$ which, ironically, led to the worst performance: The NLP is equal to the prior entropy ($1.42$). The linear estimator performed better because the regression coefficients are entirely determined by noise in the training set, \ie{} the scalar summary is a random projection of the candidate summaries. Similarly, extracting features using PLS regression is driven by noise: Here, three random orthogonal projections of candidate features were selected based on five-fold cross-validation, allowing PLS to outperform both linear and nonlinear regression. Minimizing the conditional posterior entropy and using candidate summaries without selection performed similarly and better than regression-based approaches.

MDN-compressed ABC performed as well as the gold standard likelihood-based inference ($\text{NLP} = 1.05\pm 0.01$) and better than samples drawn directly from the MDN ($\text{NLP} = 1.08\pm 0.02$), as illustrated in panel~(b) of \cref{fig:benchmark} for a particular example. While the bottleneck forces the network to compress data to an informative summary statistic, the architecture of the MDN with only two mixture components is too restrictive to approximate the true posterior well. Increasing the number of components to $k=10$ provides a better approximation with the same performance as both the likelihood-based approach and MDN-compressed ABC. Here, we deliberately restricted the architecture to illustrate that ABC with good summaries can remain competitive because it does not rely on parametric assumptions about the density.

Due to the simplicity of the benchmark problem, we can inspect the MDN and learned summary, as shown in panel~(c). The appropriate summary is obvious in retrospect: It should discriminate between data $z_{\bullet 1}$ clustered around $\pm1$ (corresponding to large absolute values of $\theta$) and data near zero or large absolute value (corresponding to small absolute values of $\theta$). The dashed line shows a polynomial approximation of the learned summary $t$ using the candidate summaries (the first three even moments of each column) as basis functions. This fit illustrates that the candidate summaries are rich enough to provide a high-quality summary in principle, but most methods struggled to extract the information. We obtained the fit by minimizing squared residuals on the interval $\parentheses{-3,3}$ weighted by the prior density. Finally, the density of the MDN, shown in panel~(d), exhibits the expected behaviour: Large summaries give rise to unimodal distributions centred at the origin, and small summaries yield bimodal posterior estimates.

The choice of compressor architecture is not unique. For example, we could have included further layers after the mean-pooling operation or used a fully-connected network throughout. However, using the mean-pooled latent features has several advantages: First, the number of compressor parameters is independent of the sample size. Second, they are unbiased estimates of the \emph{population} mean of the features independent of sample size. The architecture was motivated by the observation that the likelihood of exponential family distributions can be expressed in terms of sums (or means) of transformations of the data and preserves the i.i.d.\ structure required to connect Fisher information maximization with EPE minimization as discussed in \cref{sec:fisher}. We thus expect the learned summaries to remain informative for different sample sizes. To test this hypothesis, we repeated the analysis with $n=100$ instead of $n=10$ observations per example. The NLPs are smaller because we had access to more data: $0.68\pm0.01$ for likelihood-based inference and $0.70\pm0.01$ for MDN-compressed ABC, where the MDN was trained on the larger dataset using the same methodology as before. Running MDN-compressed ABC with the network trained on the smaller dataset yielded a NLP of $0.72\pm0.01$, \ie{} the performance is almost indistinguishable from the network trained on the larger dataset despite being exposed to an order of magnitude fewer observations. Importantly, the posterior density estimator itself cannot achieve this generalization because the model was trained on data with a fixed sample size. BayesFlow seeks to provide amortized inference even for variable sample sizes although at the cost of further simulations \citep{Radev2022BayesFlow}.

\subsection{Population genetics model\label{sec:population-genetics}}

We inferred the mutation and recombination rates of a population genetics model, a problem that has been extensively studied using ABC in general and in the context of identifying summaries in particular \citep{Joyce2008,Nunes2010,Blum2013}. Data were generated using the coalescent approach which considers the history of a sample of haplotypes, a set of DNA variations that tend to be inherited together because they are close together on the DNA strand \citep{Nordborg2019Coalescent}. We present the process in terms of the equivalent forward model because it is more accessible. Under the neutral Fisher-Wright model, diploid organisms (each having paired chromosomes) reproduce sexually in discrete generations without selection pressure. Haplotypes are subject to random mutations under an infinite-sites assumption, \ie{} the DNA sequence is sufficiently long that the probability of multiple mutations occurring at the same site is negligible. The model also allows for recombination, \ie{} the haplotype of a gamete can be a combination of parental haplotypes. We consider a finite-sites recombination model \citep{Hudson1983}, \ie{} the strands may only cross over at specific locations during meiosis. This may seem at odds with the infinite-sites mutation model, but we can think of the haplotype as a sequence of atomic segments, each comprising many bases.

We used a dataset of $10^6$ simulations from the above model generously provided by \citet{Nunes2010}. The two rate parameters were drawn from a $\mathsf{Uniform}\parentheses{0,10}$ prior. For each simulation, $50$ haplotypes with $5{,}001$ base pairs were generated. Seven candidate summaries comprised a uniform distractor and six expert summaries (such as the number of unique haplotypes or ``the frequency of the most common haplotype''~(p.~8); see \citet{Nunes2010} for details). We split the dataset into training, validation, and test sets comprising $9.89\times10^5$, $10^4$, and $10^3$ samples, respectively.

For the non-linear regression model, we used a three-layer MLP with 16, 16, and 2 hidden units without mean-pooling because the input to the neural network was a set of candidate statistics. We employed the same architecture for the compressor of the mixture density network. Instead of two Gaussian components, we used ten beta distributions rescaled to the interval $\parentheses{0,10}$ as components. We used the same architecture as in \cref{sec:benchmark} to estimate the mixture logits and the logarithm of the beta shape parameters. Both networks were trained as previously described with mini-batch size of 256.

We drew $1{,}000$ posterior samples for each example in the test set (comprising $\approx 0.1\%$ of the reference table as for the benchmark dataset). Kernel density estimates of the NLP are biased for bounded parameters because probability mass can ``leak'' out of the support \citep{Scott2015MultivariateDensityEstimation}. We used a reflection-based bias correction technique to estimate the NLP \citep{Boneva1971DensityEstimationReflection}, \ie{} each approximate posterior sample $\tilde\theta$ is reflected at the boundaries such that both tails of the kernel contribute to the density estimate. The results are shown in panel~(b) of \cref{fig:evaluation} and \cref{tbl:methods} in the appendix. Nonlinear regression and MDN-compressed ABC not only performed comparably, but the two methods also learned very similar summaries: After standardizing and aligning the summaries using a Procrustes transformation \citep{Schoenemann1966Procrustes}, they had a pointwise MSE of $M^2 = 0.20$ ($p$-value $<10^{-3}$ under a permutation test). MDN samples performed slightly, but not statistically significantly, better in terms of EPE and RMISE. The minimum CPE method performed worst in terms of NLP because it targets highly concentrated posteriors, although not necessarily near the true value. We did not implement the two-stage method of \citep{Nunes2010} due to its computational cost and similarity with posterior mean estimation approaches \citep{Fearnhead2012,Jiang2017}. Similarly, we omitted approximate sufficiency \citep{Joyce2008} because it is sensitive to the number of histogram bins chosen for density estimation and is not suitable for multidimensional parameter spaces due to the curse of dimensionality.

\subsection{Growing tree model\label{sec:tree}}

Inferring the parameters of dynamic network models is challenging, especially when only cross-sectional data are available. \citet{Cantwell2021TreePosterior} developed a pseudo-marginal Markov chain Monte Carlo (PM-MCMC) algorithm to infer the parameters of a growing tree model. Under the model \citep{Krapivsky2001GrowingNetworks}, a tree is grown from a single isolated node. At each step, a new node $j$ is added and connected to an existing node $i$ with probability $\propto k_i^\theta$, where $k_i$ is the degree of node $i$, \ie{} the number of connections it has. The parameter $\theta$ controls the strength of preferential attachment: The larger $\theta$ the more likely nodes are to connect to nodes that already have many connections. The likelihood is intractable because the history, \ie{} the order of addition of nodes, is unknown. There are combinatorially many possible histories, and evaluating the likelihood exactly is infeasible save for very small trees. The PM-MCMC algorithm estimates the marginal likelihood by sampling a set of histories consistent with the observed graph and averaging the conditional likelihood for each history \citep{Cantwell2021TreePosterior}.

Here, we employed ABC to infer the preferential attachment parameter $\theta$ and compared different methods to extract informative summaries from graph data. Synthetic data were generated by sampling the kernel parameter $\theta$ from a uniform distribution $\UniformDist\parentheses{0, 2}$ and simulating trees with $n=100$ nodes for each $\theta$. We generated a training set of $10^5$ samples; the validation and test sets both comprised $10^3$ samples.

For subset selection methods, we used one uniform distractor and four candidate summaries: The standard deviation and Gini coefficient of the degree distribution because heavy-tailed degree distributions are indicative of large $\theta$; the diameter of the network and the maximum betweenness centrality because the existence of a central hub connecting disparate parts of the tree is indicative of large $\theta$ \citep{Newman2018Networks}. For nonlinear posterior mean estimation and MDN compression, we used a two-layer graph isomorphism network (GIN) where each layer comprised a two-layer MLP with eight hidden units per layer \citep{Xu2019GraphIsomorphismNetworks}. We used the constant vector of ones as input features for the GIN because nodes are indistinguishable. Features were mean-pooled across the graph after applying the GIN, and training used 32 trees per mini-batch.

All methods, including using candidate summaries directly, significantly reduced uncertainty about the attachment parameter. As shown in panel~(c) of \cref{fig:evaluation} and cref{tbl:methods} in the appendix, the PM-MCMC algorithm had the lowest RMISE, but MDN-compressed ABC and direct sampling from the MDN performed best in terms of NLP. Similar to the benchmark experiment in \cref{sec:benchmark}, we repeated the experiment for larger trees with $n=748$ nodes. MDN compression trained on large trees performed best ($\text{NLP}=-1.67\pm0.02$), but summaries learned on trees with $n=100$ nodes generalized to larger trees with $\text{NLP}=-1.65\pm0.02$. The relative NLP performance of other methods remained unchanged, but PM-MCMC performance was sensitive to the size of the grown tree and degraded severely with $\text{NLP}=1.7\pm 0.2$, much larger than the prior entropy $\entropy\braces{\prior{\theta}}=0.69$. For superlinear preferential attachment, \ie{} $\theta > 1$, almost every new node connects to a central hub \citep{Krapivsky2001GrowingNetworks}. This phenomenon is particularly pronounced for larger graphs, and it is challenging to infer histories accurately which leads to poor inference. In this experiment, ABC with MDN-compressed summaries outperformed the dedicated (pseudo-marginal) likelihood-based approach.

\subsection{Computational cost}

\begin{table}
    \centering
    \begin{tabular}{lrrrrr}
        \toprule
        & \multicolumn{2}{c}{Benchmark} & \multicolumn{1}{c}{Coalescent} & \multicolumn{2}{c}{Growing tree} \\
        \cmidrule(lr){2-3} \cmidrule(lr){5-6}
        Step or Method & Small & Large & & Small & Large\\
        \midrule

        \multicolumn{6}{c}{\emph{Data generation}} \\
        Training set & 00:02 & 00:08 & \multicolumn{1}{c}{unknown} & 01:24 & 01:08:06\\
        Validation set & 00:02 & 00:02 & \multicolumn{1}{c}{unknown} & 00:03 & 00:48\\
        Test set & 00:02 & 00:02 & \multicolumn{1}{c}{unknown} & 00:03 & 00:47 \\

        \multicolumn{6}{c}{\emph{Training}} \\
        MDN & 04:18 & 07:13 & 05:14 & 03:28 & 21:30 \\
        Nonlinear regression & 00:36 & 02:55 & 00:59 & 03:22 & 19:54 \\

        \multicolumn{6}{c}{\emph{Inference}}\\
        Likelihood-based & 02:29 & 07:43 & \multicolumn{1}{c}{not applicable} & 07:51 & 01:11:01 \\
        Expert summaries & 00:05 & 00:13 & 00:04 & 00:52 & 25:19 \\
        MDN compression & 00:03 & 00:05 & 00:02 & 00:07 & 00:20 \\
        MDN & 00:02 & 00:02 & 00:03 & 00:02 & 00:03 \\
        Nonlinear regression & 00:03 & 00:05 & 00:02 & 00:07 & 00:19 \\
        Linear regression & 00:04 & 00:13 & 00:03 & 00:52 & 25:19 \\
        Minimum CPE & 17:04 & 14:15 & 30:38 & 5:39 & 30:17 \\
        PLS & 00:08 & 00:17 & 00:16 & 00:52 & 25:19 \\
        Prior & 00:03 & 00:03 & 00:02 & 00:02 & 00:02 \\
        \bottomrule
    \end{tabular}
    \caption{\emph{Computational costs for data generation, training of neural compressors, and inference.} Times are as hours:minutes:seconds. ``MDN'' refers to drawing samples directly from the learned mixture density network, ``likelihood-based'' refers to samples obtained with Stan for the benchmark problem and pseudo-marginal MCMC for the growing tree model, and all other methods use ABC after extracting summaries. ``Small'' and ``Large'' refer to different sample sizes for the benchmark ($n=10$ and $n=100$ samples) and growing tree ($n=100$ and $n=748$ nodes) experiments. Training times for MDN and nonlinear regression reflect a single training run. Inference times are for the complete test set of 1{,}000 examples. For methods using candidate summaries (expert summaries, linear regression, minimum CPE, and PLS), times include evaluating those summaries. Data generation times for the coalescent experiment are unknown as the dataset was provided by \citet{Nunes2010}.}
    \label{tbl:computational-cost}
\end{table}

Training neural compressors, especially MDNs, is more computationally demanding than simpler linear regression or using expert summaries directly if they are cheap to evaluate. However, as shown in \cref{tbl:computational-cost}, the relative cost of optimizing an MDN compared with non-linear regression decreases with increasing problem complexity as the neural compressor is responsible for the majority of the computational cost. For the benchmark with a simple compressor architecture, MDN training is approximately seven times slower than nonlinear regression. For the growing tree experiment with a graph neural network compressor, the additional cost is only 8\%. Training an MDN is comparable with (small benchmark) or more computationally efficient (all other experiments) than likelihood-based inference using Stan or pseudo-marginal MCMC for the growing tree experiment. Further, optimizing an MDN is a one-time expense and can extract summaries efficiently once trained. For example, computing network summaries can be costly and must be repeated for each element of the training and test sets before running ABC \citep{Raynal2023}. This is much slower than using a graph neural network compressor in our experiments: more than 25~minutes compared with only 20~seconds. Training and applying neural compressors is also more efficient than greedy subset selection using CPE minimization because ABC needs to be run multiple times for each example to iteratively select promising summaries.

\section{Discussion\label{sec:discussion}}

We have shown that five information-theoretic approaches to devising summaries are equivalent in \cref{sec:epe}. Furthermore, as shown in \cref{sec:related}, other methods can be understood as special or limiting cases of minimizing the expected posterior entropy (EPE) which should be the practitioner's choice because it is straightforward to evaluate compared with MI or KL divergence, can incorporate prior information, and is conceptually simple. We also characterized the notion of sufficient, lossless, and optimal summaries in \cref{sec:background}---distinctions that are important for developing compression algorithms and resolving misunderstandings, as discussed in \cref{sec:maxinfo}.

We compared various methods on a benchmark problem (\cref{sec:benchmark}), a population genetics model (\cref{sec:population-genetics}), and a model for growing trees (\cref{sec:tree}). Minimizing the EPE yields highly informative summaries while achieving the long-standing goal of ``find[ing] methods which do not require a preliminary subjective feature selection stage'' \citep[p.~147]{Prangle2018}. But there is no free lunch: We instead have to choose a compression and density estimation architecture. Choosing appropriate architectures can improve performance, reduce the number of simulations required \citep{Chan2018}, and even allow summaries to generalize across datasets of different sizes as demonstrated in the benchmark and growing trees experiments.

Sequential methods can reduce the computational burden of likelihood-free inference \citep{Lueckmann2017,Papamakarios2016,Chen2021MutualInfo}, but we focused on learning summaries for rejection ABC for two reasons: First, we wanted to isolate the effect of summary selection without introducing confounders. We omitted regression adjustment for ABC samples \citep{Beaumont2002ABC} for the same reason. Second, learning global summaries allows for amortized inference because we do not need to retrain models for each example. Investigating the interaction between sequential methods and learning summaries could shed light on how different aspects of the data inform parameters in different regions of the parameter space, as illustrated in \cref{fig:piecewise}.

The summaries of the mixture density networks in \cref{sec:experiments} can have arbitrary scales which can be problematic for ABC. We standardized summaries after extraction to mitigate this problem, but metric learning approaches could further improve ABC with MDN-compressed summaries \citep{GonzalezVanegas2019}. Investigating the impact of model misspecification on ABC is an active area of research \citep{Frazier2020}, and comparing the robustness of different methods should be considered in future work.

Neural density estimation is a powerful tool for likelihood-free inference, ``but there is no uniformly best algorithm'' \citep[p.~1]{Lueckmann2021}. ABC remains a compelling approach because of its theoretical properties, and it can produce high-fidelity posteriors, especially when low-dimensional but rich summaries can be extracted from complex data.

\section*{Acknowledgement}

We thank Matthew Nunes for sharing simulations and John Platig for feedback on the population genetics problem.

\section*{Statements and Declarations}

\subsection*{Funding}

This work was supported by the U.S. National Institutes of Health under Grant~R01AI138901 and~R35CA220523.

\subsection*{Competing Interests}

The authors have no relevant financial or non-financial interests to disclose.

\subsection*{Code and Data Availability}

The population genetics simulations were generously provided by Matthew Nunes and are available at \url{https://github.com/onnela-lab/coaloracle}. All code and other simulations are available at \url{https://github.com/onnela-lab/summaries}.

\bibliography{main}

@article{Nunes2010,
	author = {Nunes, Matthew A. and Balding, David J.},
	doi = {10.2202/1544-6115.1576},
	journal = {Stat. Appl. Genet. Mol. Biol.},
	number = {1},
	title = {On Optimal Selection of Summary Statistics for Approximate {Bayesian} Computation},
	volume = {9},
	year = {2010},
}

@article{Heavens2000,
	author = {Heavens, Alan F. and Jimenez, Raul and Lahav, Ofer},
	doi = {10.1046/j.1365-8711.2000.03692.x},
	journal = {Mon. Not. R. Astron. Soc.},
	month = {10},
	number = {4},
	pages = {965--972},
	title = {Massive lossless data compression and multiple parameter estimation from galaxy spectra},
	volume = {317},
	year = {2000},
}

@inproceedings{Nirwan2019,
	author = {Nirwan, Rajbir and Bertschinger, Nils},
	booktitle = {International Conference on Machine Learning},
	pages = {4820--4828},
	title = {Rotation Invariant {Householder} Parameterization for {Bayesian} {PCA}},
	volume = {97},
	year = {2019},
}

@article{Hoff2002,
	author = {Hoff, Peter D. and Raftery, Adrian E. and Handcock, Mark S.},
	doi = {10.1198/016214502388618906},
	journal = {J. Am. Stat. Assoc.},
	number = {460},
	pages = {1090--1098},
	title = {Latent Space Approaches to Social Network Analysis},
	volume = {97},
	year = 2002,
}

@article{Stephens2000,
	author = {Stephens, Matthew},
	doi = {10.1111/1467-9868.00265},
	journal = {J. R. Stat. Soc. Ser. B Stat. Methodol.},
	number = {4},
	pages = {795--809},
	title = {Dealing with label switching in mixture models},
	volume = {62},
	year = {2000},
}

@article{Blum2013,
	author = {Blum, Michael G. B. and Nunes, Matthew A. and Prangle, Dennis and Sisson, Scott A.},
	doi = {10.1214/12-STS406},
	journal = {Stat. Sci.},
	number = {2},
	pages = {189--208},
	title = {A Comparative Review of Dimension Reduction Methods in Approximate {Bayesian} Computation},
	volume = {28},
	year = {2013},
}

@article{Blum2010,
	author = {Blum, Michael G. B. and Fran{\c c}ois, Olivier},
	doi = {10.1007/s11222-009-9116-0},
	journal = {Stat. Comput.},
	number = {1},
	pages = {63--73},
	title = {Non-linear regression models for Approximate {Bayesian} Computation},
	volume = {20},
	year = {2010},
}

@incollection{Prangle2018,
	author = {Prangle, Dennis},
	booktitle = {Handbook of Approximate {Bayesian} Computation},
	editor = {Sisson, Scott A. and Fan, Yanan and Beaumount, Mark},
	publisher = {Chapman \& Hall/CRC},
	title = {Summary statistics},
	year = {2018},
    pages = {125--152},
    address = {Philadelphia},
}

@inproceedings{Chan2018,
	author = {Chan, Jeffrey and Perrone, Valerio and Spence, Jeffrey P. and Jenkins, Paul A. and Mathieson, Sara and Song, Yun S.},
	booktitle = {Advances in Neural Information Processing Systems},
	pages = {8603--8614},
	title = {A Likelihood-Free Inference Framework for Population Genetic Data Using Exchangeable Neural Networks},
	volume = {32},
	year = {2018},
}

@article{Joyce2008,
	author = {Joyce, Paul and Marjoram, Paul},
	doi = {10.2202/1544-6115.1389},
	journal = {Stat. Appl. Genet. Mol. Biol.},
	number = {1},
	title = {Approximately Sufficient Statistics and {Bayesian} Computation},
	volume = {7},
	year = {2008},
}

@inproceedings{Lueckmann2021,
	author = {Lueckmann, Jan-Matthis and Boelts, Jan and Greenberg, David and Goncalves, Pedro and Macke, Jakob},
	booktitle = {International Conference on Artificial Intelligence and Statistics},
	pages = {343--351},
	title = {Benchmarking Simulation-Based Inference},
	volume = {130},
	year = {2021},
}

@article{Raynal2023,
	author = {Raynal, Louis and Hoffmann, Till and Onnela, Jukka-Pekka},
	doi = {10.1080/10618600.2022.2151453},
	journal = {J. Comput. Graph. Stat.},
	number = {0},
	pages = {1--10},
	title = {Cost-based Feature Selection for Network Model Choice},
	volume = {0},
	year = {2023},
}

@article{Yang2019,
	author = {Yang, Yingrui and McKhann, Ashley and Chen, Sixing and Harling, Guy and Onnela, Jukka-Pekka},
	doi = {10.1016/j.epidem.2019.03.002},
	journal = {Epidemics},
	pages = {115--122},
	title = {Efficient vaccination strategies for epidemic control using network information},
	volume = {27},
	year = {2019},
}

@article{Frazier2020,
	doi = {10.1111/rssb.12356},
	year = 2020,
	volume = {82},
	number = {2},
	pages = {421--444},
	author = {Frazier, David T. and Robert, Christian P. and Rousseau, Judith},
	title = {Model Misspecification in Approximate {Bayesian} Computation: Consequences and Diagnostics},
	journal = {J. R. Stat. Soc. Ser. B Stat. Methodol.}
}

@inproceedings{Paszke2019,
	author = {Paszke, Adam and Gross, Sam and Massa, Francisco and Lerer, Adam and Bradbury, James and Chanan, Gregory and Killeen, Trevor and Lin, Zeming and Gimelshein, Natalia and Antiga, Luca and Desmaison, Alban and Kopf, Andreas and Yang, Edward and DeVito, Zachary and Raison, Martin and Tejani, Alykhan and Chilamkurthy, Sasank and Steiner, Benoit and Fang, Lu and Bai, Junjie and Chintala, Soumith},
	booktitle = {Advances in Neural Information Processing Systems},
	pages = {8024--8035},
	title = {{PyTorch}: An Imperative Style, High-Performance Deep Learning Library},
	volume = {32},
	year = {2019},
}

@article{Koopman1936,
	author = {Koopman, Bernard O.},
	doi = {10.1090/S0002-9947-1936-1501854-3},
	journal = {Trans. Am. Math. Soc.},
	number = {3},
	pages = {399--409},
	title = {On Distributions Admitting a Sufficient Statistic},
	volume = {39},
	year = {1936},
}

@incollection{Nordborg2019Coalescent,
	author = {Nordborg, Magnus},
	booktitle = {Handbook of Statistical Genomics},
	editor = {Balding, David J. and Moltke, Ida and Marioni, John},
	publisher = {Wiley},
	title = {Coalescent theory},
	volume = {1},
	year = {2019},
    doi = {10.1002/9781119487845.ch5},
    pages = {145--175},
    address = {Hoboken},
}

@article{Charnock2018,
	author = {Charnock, Tom and Lavaux, Guilhem and Wandelt, Benjamin D.},
	doi = {10.1103/PhysRevD.97.083004},
	journal = {Phys. Rev. D},
	number = {8},
	pages = {083004},
	title = {Automatic physical inference with information maximizing neural networks},
	volume = {97},
	year = {2018},
}

@article{Grassmann2024ProteinInteractions,
	author = {Grassmann, Greta and Miotto, Mattia and Desantis, Fausta and Di Rienzo, Lorenzo and Tartaglia, Gian Gaetano and Pastore, Annalisa and Ruocco, Giancarlo and Monti, Michele and Milanetti, Edoardo},
	doi = {10.1021/acs.chemrev.3c00550},
	journal = {Chem. Rev.},
	number = {7},
	pages = {3932--3977},
	title = {Computational Approaches to Predict Protein--Protein Interactions in Crowded Cellular Environments},
	volume = {124},
	year = {2024},
}

@article{Alsing2018,
	author = {Alsing, Justin and Wandelt, Benjamin},
	doi = {10.1093/mnrasl/sly029},
	journal = {Mon. Not. R. Astron. Soc. Lett.},
	number = {1},
	pages = {L60-L64},
	title = {Generalized massive optimal data compression},
	volume = {476},
	year = {2018},
}

@article{Fearnhead2012,
	author = {Fearnhead, Paul and Prangle, Dennis},
	doi = {10.1111/j.1467-9868.2011.01010.x},
	journal = {J. R. Stat. Soc. Ser. B Stat. Methodol.},
	number = {3},
	pages = {419--474},
	title = {Constructing summary statistics for approximate {Bayesian} computation: semi-automatic approximate {Bayesian} computation},
	volume = {74},
	year = {2012},
}

@article{Jiang2017,
	author = {Jiang, Bai and Wu, Tung-Yu and Zheng, Charles and Wong, Wing H.},
	doi = {10.5705/ss.202015.0340},
	journal = {Stat. Sin.},
	pages = {1595--1618},
	title = {Learning summary statistics for approximate {Bayesian} computation via deep neural network},
	volume = {27},
	year = {2017},
}

@inproceedings{Papamakarios2016,
	author = {Papamakarios, George and Murray, Iain},
	booktitle = {Advances in Neural Information Processing Systems},
	title = {Fast $\epsilon$-free Inference of Simulation Models with {Bayesian} Conditional Density Estimation},
	volume = {29},
	year = {2016},
}

@article{Papamakarios2021,
	author = {Papamakarios, George and Nalisnick, Eric and Rezende, Danilo Jimenez and Mohamed, Shakir and Lakshminarayanan, Balaji},
	journal = {J. Mach. Learn. Res.},
	number = {57},
	pages = {1--64},
	title = {Normalizing Flows for Probabilistic Modeling and Inference},
	volume = {22},
	year = {2021},
}

@inproceedings{Lueckmann2017,
	author = {Lueckmann, Jan-Matthis and Goncalves, Pedro J. and Bassetto, Giacomo and \"{O}cal, Kaan and Nonnenmacher, Marcel and Macke, Jakob H.},
	booktitle = {Advances in Neural Information Processing Systems},
	title = {Flexible statistical inference for mechanistic models of neural dynamics},
	volume = {30},
	year = {2017},
}

@article{Jeffrey2020NeuralSummaries,
    title = "Likelihood-free inference with neural compression of {DES} {SV} weak lensing map statistics",
    journal = "Mon. Not. R. Astron. Soc.",
    volume = "501",
    pages = "954--969",
    doi = "10.1093/mnras/staa3594",
    number = "1",
    author = "Jeffrey, Niall and Alsing, Justin and Lanusse, François",
    year = "2020",
}

@article{Krapivsky2001GrowingNetworks,
    title = "Organization of growing random networks",
    doi = "10.1103/PhysRevE.63.066123",
    volume = "63",
    pages = "66123",
    journal = "Phys. Rev. E",
    number = "6",
    author = "Krapivsky, Pavel L. and Redner, Sidney",
    year = "2001",
}

@incollection{Vaart1998,
	author = {{van der Vaart}, Aad W.},
	booktitle = {Asymptotic statistics},
	doi = {10.1017/CBO9780511802256},
	pages = {140--146},
	publisher = {Cambridge University Press},
	title = {{Bernstein--von Mises} Theorem},
	year = {1998},
    address = {Cambridge},
}

@article{Barnes2012,
	author = {Barnes, Chris P. and Filippi, Sarah and Stumpf, Michael P. H. and Thorne, Thomas},
	doi = {10.1007/s11222-012-9335-7},
	journal = {Stat. Comput.},
	number = {6},
	pages = {1181--1197},
	title = {Considerate approaches to constructing summary statistics for {ABC} model selection},
	volume = {22},
	year = {2012},
}

@article{Itti2009,
	author = {Itti, Laurent and Baldi, Pierre},
	doi = {10.1016/j.visres.2008.09.007},
	journal = {Vision Res.},
	number = {10},
	pages = {1295--1306},
	title = {Bayesian surprise attracts human attention},
	volume = {49},
	year = {2009},
}

@article{Singh2003,
	author = {Singh, Harshinder and Misra, Neeraj and Hnizdo, Vladimir and Fedorowicz, Adam and Demchuk, Eugene},
	doi = {10.1080/01966324.2003.10737616},
	journal = {Am. J. Math. Manag. Sci.},
	number = {3--4},
	pages = {301--321},
	title = {Nearest Neighbor Estimates of Entropy},
	volume = {23},
	year = {2003},
}

@article{Carpenter2017,
	author = {Carpenter, Bob and Gelman, Andrew and Hoffman, Matthew D. and Lee, Daniel and Goodrich, Ben and Betancourt, Michael and Brubaker, Marcus and Guo, Jiqiang and Li, Peter and Riddell, Allen},
	doi = {10.18637/jss.v076.i01},
	journal = {J. Stat. Softw.},
	number = {1},
	pages = {1--32},
	title = {Stan: A Probabilistic Programming Language},
	volume = {76},
	year = {2017},
}

@article{Hudson1983,
	author = {Hudson, Richard R.},
	doi = {10.1016/0040-5809(83)90013-8},
	journal = {Theor. Popul. Biol.},
	number = {2},
	pages = {183--201},
	title = {Properties of a neutral allele model with intragenic recombination},
	volume = {23},
	year = {1983},
}

@inproceedings{Kingma2015,
	author = {Kingma, Diederik P. and Ba, Jimmy},
	booktitle = {International Conference on Learning Representations},
	title = {Adam: A Method for Stochastic Optimization},
	year = {2015},
}

@techreport{Bishop1994,
	author = {Bishop, Cristopher M.},
	institution = {Aston University},
	number = {NCRG/94/004},
	title = {Mixture density networks},
	year = {1994},
}

@article{Beaumont2019,
	author = {Beaumont, Mark A.},
	doi = {10.1146/annurev-statistics-030718-105212},
	journal = {Annu. Rev. Stat. Appl.},
	number = {1},
	pages = {379--403},
	title = {Approximate {Bayesian} Computation},
	volume = {6},
	year = {2019},
}

@article{Prangle2014,
	author = {Prangle, Dennis and Fearnhead, Paul and Cox, Murray P. and Biggs, Patrick J. and French, Nigel P.},
	doi = {10.1515/sagmb-2013-0012},
	journal = {Stat. Appl. Genet. Mol. Biol.},
	number = {1},
	pages = {67--82},
	title = {Semi-automatic selection of summary statistics for {ABC} model choice},
	volume = {13},
	year = {2014},
}

@article{Merten2019,
	author = {Merten, Julian and Giocoli, Carlo and Baldi, Marco and Meneghetti, Massimo and Peel, Austin and Lalande, Florian and Starck, Jean-Luc and Pettorino, Valeria},
	doi = {10.1093/mnras/stz972},
	journal = {Mon. Not. R. Astron. Soc.},
	number = {1},
	pages = {104--122},
	title = {On the dissection of degenerate cosmologies with machine learning},
	volume = {487},
	year = {2019},
}

@article{Cai2022,
	author = {Cai, Yuhang and Lim, Lek-Heng},
	doi = {10.1109/TIT.2022.3148923},
	journal = {IEEE Trans. Inf. Theory},
	title = {Distances between probability distributions of different dimensions},
	year = {2022},
}

@article{Kramer1991Autoencoder,
    title = "Nonlinear principal component analysis using autoassociative neural networks",
    journal = "AIChE J.",
    volume = "37",
    pages = "233--243",
    doi = "10.1002/aic.690370209",
    number = "2",
    author = "Kramer, Mark A.",
    year = "1991",
}

@article{Boneva1971DensityEstimationReflection,
    title = "Spline Transformations: Three New Diagnostic Aids for the Statistical Data-Analyst",
    journal = "J. R. Stat. Soc. Ser. B Stat. Methodol.",
    volume = "33",
    pages = "1--37",
    doi = "10.1111/j.2517-6161.1971.tb00855.x",
    number = "1",
    author = "Boneva, Liliana I. and Kendall, David and Stefanov, Ivan",
    year = "1971",
}

@article{GonzalezVanegas2019,
    title = "{AKL-ABC}: An Automatic Approximate {Bayesian} Computation Approach Based on Kernel Learning",
    journal = "Entropy",
    volume = "21",
    pages = "932",
    doi = "10.3390/e21100932",
    number = "10",
    author = {Gonz{\'a}lez-Vanegas, Wilson and {\'A}lvarez-Meza, Andr{\'e}s and Hern{\'a}ndez-Muriel, Jos{\'e} and Orozco-Guti{\'e}rrez, {\'A}lvaro},
    year = "2019",
}

@article{Schoenemann1966Procrustes,
    title = "A generalized solution of the orthogonal {Procrustes} problem",
    journal = "Psychometrika",
    volume = "31",
    pages = "1--10",
    doi = "10.1007/bf02289451",
    number = "1",
    author = {Sch{\"o}nemann, Peter H.},
    year = "1966",
}

@book{Scott2015MultivariateDensityEstimation,
    title = "Multivariate Density Estimation",
    publisher = "Wiley",
    doi = "10.1002/9781118575574",
    author = "Scott, David W.",
    year = "2015",
    address = "Hoboken",
}

@article{Radev2022BayesFlow,
    title = "{BayesFlow}: Learning Complex Stochastic Models With Invertible Neural Networks",
    journal = {IEEE Trans. Neural Netw. Learn. Syst.},
    volume = "33",
    pages = "1452--1466",
    doi = "10.1109/tnnls.2020.3042395",
    number = "4",
    author = {Radev, Stefan T. and Mertens, Ulf K. and Voss, Andreas and Ardizzone, Lynton and Kothe, Ullrich},
    year = "2022",
}

@inproceedings{Xu2019GraphIsomorphismNetworks,
    title = "How Powerful are Graph Neural Networks?",
    booktitle = "International Conference on Learning Representations",
    volume = "7",
    author = "Xu, Keyulu and Hu, Weihua and Leskovec, Jure and Jegelka, Stefanie",
    year = "2019",
}

@article{Beaumont2002ABC,
    title = "Approximate {Bayesian} Computation in Population Genetics",
    journal = "Genetics",
    volume = "162",
    pages = "2025--2035",
    doi = "10.1093/genetics/162.4.2025",
    number = "4",
    author = "Beaumont, Mark A. and Zhang, Wenyang and Balding, David J.",
    year = "2002",
}

@article{Wegmann2009PartialLeastSquares,
    title = "Efficient Approximate {Bayesian} Computation Coupled With {Markov} Chain {Monte} {Carlo} Without Likelihood",
    journal = "Genetics",
    volume = "182",
    pages = "1207--1218",
    doi = "10.1534/genetics.109.102509",
    number = "4",
    author = "Wegmann, Daniel and Leuenberger, Christoph and Excoffier, Laurent",
    year = "2009",
}

@article{Cantwell2021TreePosterior,
    title = "Inference, Model Selection, and the Combinatorics of Growing Trees",
    journal = "Phys. Rev. Lett.",
    volume = "126",
    doi = "10.1103/physrevlett.126.038301",
    number = "3",
    author = "Cantwell, George T. and St-Onge, Guillaume and Young, Jean-Gabriel",
    year = "2021",
}

@book{Bishop2006,
	author = {Bishop, Cristopher M.},
	publisher = {Springer},
	title = {Pattern recognition and machine learning},
	year = {2006},
    address = {New York},
}

@article{Sobkowicz2012,
	author = {Sobkowicz, Pawel and Kaschesky, Michael and Bouchard, Guillaume},
	doi = {10.1016/j.giq.2012.06.005},
	journal = {Gov. Inf. Q.},
	number = {4},
	pages = {470--479},
	title = {Opinion mining in social media: Modeling, simulating, and forecasting political opinions in the web},
	volume = {29},
	year = {2012},
}

@article{Burr2013Calibration,
    title = {Selecting Summary Statistics in Approximate {Bayesian} Computation for Calibrating Stochastic Models},
    author = {Burr, Tom and Skurikhin, Alexei},
    year = {2013},
    journal = {BioMed Res. Int.},
    pages = {210646},
    doi = {10.1155/2013/210646},
}

@article{Aeschbacher2012,
	author = {Aeschbacher, Simon and Beaumont, Mark A. and Futschik, Andreas},
	doi = {10.1534/genetics.112.143164},
	journal = {Genetics},
	month = {11},
	number = {3},
	pages = {1027--1047},
	title = {A Novel Approach for Choosing Summary Statistics in Approximate {Bayesian} Computation},
	volume = {192},
	year = {2012},
}

@book{Newman2018Networks,
    title = "Networks",
    publisher = "Oxford University Press",
    doi = "10.1093/oso/9780198805090.001.0001",
    author = "Newman, Mark",
    year = "2018",
    address = "Oxford",
}

@inproceedings{Chen2021MutualInfo,
	author = {Chen, Yanzhi and Zhang, Dinghuai and Gutmann, Michael U. and Courville, Aaron and Zhu, Zhanxing},
	booktitle = {International Conference on Learning Representations},
	title = {Neural Approximate Sufficient Statistics for Implicit Models},
	year = {2021},
}

@article{Fluri2021,
	author = {Fluri, Janis and Kacprzak, Tomasz and Refregier, Alexandre and Lucchi, Aurelien and Hofmann, Thomas},
	doi = {10.1103/PhysRevD.104.123526},
	journal = {Phys. Rev. D},
	number = {12},
	pages = {123526},
	title = {Cosmological parameter estimation and inference using deep summaries},
	volume = {104},
	year = {2021},
}

\appendix

\section{EPE and expected Kullback-Leibler divergence\label{app:epe-kl-relationship}}

The difference between the EPE given only the summaries $t$ and the EPE given the underlying data $z$ is
\[
    \epe{}-\E{\entropy\braces{\posterior{\theta\mid z}}}_{z\dist\proba{z}}
    =\int\diff{z}\diff{\theta}\proba{z,\theta}\log\posterior{\theta\mid z}-\int\diff{t}\diff{\theta}\proba{t,\theta}\log\posterior{\theta\mid t}.
\]
Changing variables of integration to $z$ in the second term and combining integrals yields
\[
    \epe{}-\E{\entropy\braces{\posterior{\theta\mid z}}}_{z\dist\proba{z}}
    =\int\diff{z}\diff{\theta}\proba{z,\theta}\brackets{\log\posterior{\theta\mid z}-\log\posterior{\theta\mid t(z)}}.
\]
Splitting the joint distribution $\proba{z,\theta}$ into conditionals $\proba{z}\posterior{\theta\mid z}$ and combining the logarithms gives the expression in the main text:
\[
    \epe{}-\E{\entropy\braces{\posterior{\theta\mid z}}}_{z\dist\proba{z}}=\int\diff{z}\proba{z}\int\diff{\theta}\posterior{\theta\mid z}\log\parentheses{\frac{\posterior{\theta\mid z}}{\posterior{\theta\mid t}}}.
\]

\section{Proposition of sufficiency by Chen et al.\label{app:chen}}

\citet{Chen2021MutualInfo} made the following proposition (see p.~2); we have adapted the notation for consistency with the main text.

\begin{prop}\label{prop:chen}
    Let $\theta\dist\prior{\theta}$, $z\dist\likelihood{z\mid\theta}$, and $t\in\mathcal{T}$ be a deterministic function. Then $t=t(z)$ is a sufficient statistic for $\likelihood{z\mid\theta}$ if and only if
    \begin{equation}
        t=\argmax_{t'\in\mathcal{T}} I\braces{\theta, t'(z)},\label{eq:chen-proposition}
    \end{equation}
    where $I\braces{\cdot,\cdot}$ denotes the mutual information between two random variables.
\end{prop}

However, as discussed in \cref{sec:background}, the existence of sufficient statistics is a property of the likelihood--not the approach used to compress the data. The statistic in \cref{eq:chen-proposition} is sufficient if and only if the likelihood belongs to the exponential family. If the likelihood does not belong to the exponential family, the statistic is optimal in the sense that it minimises \cref{eq:optimal-defn} with loss functional being the expected Kullback-Leibler divergence, as discussed in \cref{sec:epe}.

They subsequently argued that ``the sufficiency of the learned statistics is insensitive to the choice of $\prior{\theta}$'' (p.~4) and that the ``approach differs from [other] methods as it is globally sufficient for all $\theta$'' (p.~6). As demonstrated in \cref{sec:maxinfo}, this assertion does not hold in general. But it is true if the likelihood belongs to the exponential family and $\mathcal{T}$ includes the sufficient statistics. The proof proposed by \citet{Chen2021MutualInfo} does not hold because it assumes the conclusion~(p.~13).

\section{Likelihood-based inference for the benchmark problem\label{app:benchmark-stan}}

We used the likelihood-based inference framework Stan~\citep{Carpenter2017} to draw samples from the true posterior for the benchmark problem in \cref{sec:benchmark}. The algorithm uses the geometry of the posterior to efficiently draw samples from it. However, complex geometries make exploring the posterior difficult, and a change of variables can be an effective means to improve sampling. For the problem at hand, we made the change of variables $\phi=\tanh\theta$ such that the likelihood becomes
\[
    y_i\mid\phi \dist \frac{1}{2}\sum_{u\in\braces{-1,1}}\NormalDist\left(u\times \phi, 1 - \phi^2\right),
\]
and we restricted $0\leq \phi < 1$ because the posterior is symmetric and exploring one mode is sufficient. The posterior is thus
\[
    \proba{\phi\mid y} \propto \frac{1}{1-\phi^2}\times\proba{\theta=\tanh^{-1}\phi}\times\prod_{i=1}^n \proba{y_i\mid\phi},
\]
where the first term is the Jacobian accounting for the change of variables. We obtained samples of $\theta$ by applying the inverse $\tanh$ transform and randomly reversing the sign with probability $0.5$. We set the target acceptance probability to $0.99$ to ensure that there are no divergent transitions which indicate numerical instabilities~\citep{Carpenter2017}.

\begin{sidewaystable}
{\small\begin{tabular}{lrrrrrr}
\toprule
& \multicolumn{2}{c}{Benchmark} & \multicolumn{2}{c}{Coalescent} & \multicolumn{2}{c}{Growing tree} \\
\cmidrule(lr){2-3}\cmidrule(lr){4-5}\cmidrule(lr){6-7}
Method & \multicolumn{1}{c}{NLP} & \multicolumn{1}{c}{RMISE} & \multicolumn{1}{c}{NLP} & \multicolumn{1}{c}{RMISE} & \multicolumn{1}{c}{NLP} & \multicolumn{1}{c}{RMISE} \\
\midrule
Likelihood-based & $\mathbf{1.05 \pm 0.01}$ & $1.24 \pm 0.02$ & \multicolumn{2}{c}{not applicable} & $-0.666 \pm 0.028$ & $\mathbf{0.165 \pm 0.002}$ \\
Expert summaries & $1.12 \pm 0.01$ & $1.25 \pm 0.02$ & $3.685 \pm 0.025$ & $4.05 \pm 0.03$ & $-0.579 \pm 0.014$ & $0.208 \pm 0.002$ \\
MDN compression & $\mathbf{1.05 \pm 0.01}$ & $1.24 \pm 0.02$ & $\mathbf{3.601 \pm 0.028}$ & $\mathbf{3.82 \pm 0.03}$ & $\mathbf{-0.730 \pm 0.021}$ & $\mathbf{0.167 \pm 0.002}$ \\
MDN & $1.08 \pm 0.02$ & $1.24 \pm 0.02$ & $\mathbf{3.595 \pm 0.028}$ & $\mathbf{3.82 \pm 0.03}$ & $\mathbf{-0.727 \pm 0.021}$ & $\mathbf{0.167 \pm 0.002}$ \\
Nonlinear regression & $1.43 \pm 0.02$ & $1.35 \pm 0.01$ & $\mathbf{3.604 \pm 0.028}$ & $\mathbf{3.82 \pm 0.03}$ & $\mathbf{-0.702 \pm 0.021}$ & $0.172 \pm 0.002$ \\
Linear regression & $1.32 \pm 0.02$ & $1.30 \pm 0.02$ & $3.677 \pm 0.029$ & $3.98 \pm 0.03$ & $-0.685 \pm 0.021$ & $0.174 \pm 0.002$ \\
Minimum CPE & $1.13 \pm 0.02$ & $\mathbf{1.20 \pm 0.02}$ & $3.734 \pm 0.033$ & $3.99 \pm 0.03$ & $\mathbf{-0.700 \pm 0.022}$ & $0.169 \pm 0.002$ \\
PLS & $1.15 \pm 0.02$ & $1.26 \pm 0.02$ & $3.712 \pm 0.024$ & $4.05 \pm 0.03$ & $-0.525 \pm 0.013$ & $0.225 \pm 0.002$ \\
Prior & $1.44 \pm 0.03$ & $1.36 \pm 0.01$ & $4.621 \pm 0.003$ & $5.69 \pm 0.03$ & $0.696 \pm 0.002$ & $0.801 \pm 0.006$ \\
\bottomrule
\end{tabular}}
\caption{\emph{The quality of summaries has a significant impact on the fidelity of posteriors.} The table reports the negative log probability loss (NLP) and root mean integrated squared error (RMISE) for combinations of methods and experiments. ``MDN'' refers to directly sampling from the learned mixture density network, ``likelihood-based'' refers to samples obtained with Stan for the benchmark problem, and all other methods use ABC after extracting summaries. Reported errors are standard errors, and methods that are within one standard error of the best method are highlighted in bold. See \cref{sec:benchmark,sec:population-genetics,sec:tree} for details on the benchmark, coalescent, and growing tree experiments, respectively.}
    \label{tbl:methods}
\end{sidewaystable}

\end{document}